\documentclass[11pt,a4paper,preprint]{IEEEtran}
\usepackage{graphicx,amsmath,amssymb}
\usepackage{hyperref} % 保留你原来所有自定义宏包

\usepackage{multirow}
\usepackage{graphicx}
\usepackage{threeparttable}
\usepackage{enumitem}
\usepackage{mathrsfs}
\usepackage{color}
\graphicspath{{figures/}}
\usepackage{algorithm,algorithmic}
\usepackage{bm}
\usepackage{amsmath,amsfonts}
\usepackage{algorithmic}
\usepackage{algorithm}
\usepackage{bm}
\usepackage{color, xcolor}
\usepackage{array}
\usepackage[caption=false,font=normalsize,labelfont=sf,textfont=sf]{subfig}
\usepackage{textcomp}
\usepackage{stfloats}
\usepackage{url}
\usepackage{verbatim}
\usepackage{graphicx}
\usepackage{cite}
\usepackage{amssymb} 
\usepackage{multirow}
\usepackage{makecell}
\usepackage{framed} 
\usepackage{setspace}
\title {Efficiency Meets Reliability: Enhanced Generalized Interleaved Transform for Random Multiplexing}

\author{
Ming~Wang,
Shufeng~Li,~\IEEEmembership{Member,~IEEE},
Lei~Liu,~\IEEEmembership{Senior Member,~IEEE},
Yao~Ge,~\IEEEmembership{Member,~IEEE},
and
Yuhao~Chi,~\IEEEmembership{Senior Member,~IEEE}
\thanks{Received January 15, 2026; accepted May 13, 2026. This paper has been accepted for publication in \textit{Chinese Journal of Electronics}, 2026. DOI will be added upon formal publication.(\emph{Corresponding author: Lei Liu.})}
\thanks{Ming Wang and Shufeng Li are with the State Key Laboratory of Media Convergence and Communication, School of Information and Communication Engineering, Communication University of China, Beijing 100024, China (e-mail: \{wangming12136, lishufeng\}@cuc.edu.cn).}
\thanks{Lei Liu is with the Zhejiang Provincial Key Laboratory of Multi-Modal Communication Networks and Intelligent Information Processing, College of Information Science and Electronic Engineering, Zhejiang University, Hangzhou 310007, China (e-mail: lei$\_$liu@zju.edu.cn).}
\thanks{Yao Ge is with the AUMOVIO-NTU Corporate Laboratory, Nanyang Technological University, Singapore 639798 (e-mail: yao.ge@ntu.edu.sg).}
\thanks{Yuhao Chi is with the State Key Laboratory of Integrated Services Networks, School of Telecommunications Engineering, Xidian University, Xi’an 710071, China (e-mail: yhchi@xidian.edu.cn).}
}

\begin{document}
\maketitle

\begin{abstract}
To meet the demands of 6G wireless systems operating in high-mobility scenarios, this paper presents a design of a random multiplexing (RM) communication system that is both storage-efficient and highly reliable. In principle, RM with cross‑domain memory approximate message passing (CD‑MAMP) can achieve replica maximum \emph{a posteriori} (MAP)-optimal performance by constructing a fully dense equivalent channel matrix. However, its practical implementation is hindered by the large storage overhead of conventional interleavers and by performance degradation in severely ill-conditioned channels, which existing related work (focusing on interleaving and transform designs) fails to address simultaneously. To overcome these issues, we develop a storage‑efficient and highly reliable system that integrates RM with CD-MAMP, referred to as RM-MAMP. Specifically, we propose a Logistic chaotic mapping interleaver with a quantitative parameter-selection criterion, and a dual-stage high-order permutation polynomial interleaver, both of which achieve nearly identical bit-error-rate (BER) as fully random interleavers while reducing the interleaver storage from ${\cal O}(N)$ to ${\cal O}(1)$ and significantly lowering interleaver signaling overhead. We further propose a highly reliable interleaved transform framework, comprising an interleaved phase perturbation transform and a multi-layer interleaved coupled transform, to enhance the incoherence and diversity of the equivalent channel matrix. Simulation results show that the proposed storage‑efficient interleavers maintain BER performance comparable to fully random interleavers, while the highly reliable transforms provide over 4 dB gain in severely time-varying channels, confirming the dual benefits of reduced storage overhead and improved robustness for the enhanced RM-MAMP system.
\end{abstract}


\section{Introduction}
\label{sec1}
Emerging 6G wireless communication systems operate in higher frequency bands, exploit wider bandwidths, and face more challenging high-mobility scenarios \cite{6G,CJE-6G}. Under such conditions, ensuring high reliability and efficiency within limited spectrum, hardware, and storage constraints is essential for the design of a communication system \cite{6G-2,CJE-6G2}. In particular, high Doppler and rapidly time-varying channel conditions impose stricter requirements on reliable and efficient data transmission and signal detection. Accordingly, from a system-level perspective, designing a joint transmission–detection solution that delivers both storage efficiency and high reliability in severely time-varying channels is a critical research direction for next-generation wireless communications. In multicarrier communication systems, modulation and detection are tightly coupled: the modulation scheme determines the structure of the equivalent channel at the receiver, thereby affecting both the choice of detector and its performance. Orthogonal frequency division multiplexing (OFDM) is a prevalent multicarrier scheme in wireless communications due to its robustness against multipath fading \cite{2008OFDM}. However, in high-mobility scenarios, OFDM suffers from severe inter-carrier interference, highlighting the urgent need for novel multicarrier modulation schemes \cite{2009OFDM}. Furthermore, iterative detectors based on the approximate message passing (AMP) algorithm have recently attracted considerable interest as innovative signal detection methods \cite{2009AMP}. Nevertheless, AMP relies on the assumption of independent and identically distributed (IID) channels and may diverge or exhibit degraded performance under correlated channel conditions \cite{AMPconverge}. Accordingly, a range of AMP-type algorithms has emerged, and combining these methods with diverse multicarrier modulation schemes has become a major research focus.
\begin{table*}[htbp]
\centering
\caption{Comparison of Representative Modulation Schemes and Detection Algorithms.}
\label{tab:M_D_Com}
\scalebox{0.76}{
\begin{threeparttable} 
\begin{tabular}{|c|c|c|c|c|}
\hline
{Modulation} & {\begin{tabular}[c]{@{}c@{}}Equivalent Channel \\ Structure\end{tabular}} & {\begin{tabular}[c]{@{}c@{}}Representative \\ Detection Algorithm\end{tabular}} & {Main Complexity Order}\tnote{1} & {Core Limitations} \\ \hline
OFDM & Diagonal (static) & \begin{tabular}[c]{@{}c@{}}LMMSE\end{tabular} & $\mathcal{O}(N^3)$ & Severe inter-carrier interference under high Doppler \\ \hline
\multirow{4}{*}{OTFS} & \multirow{4}{*}{Approximately sparse} & CD-OAMP & $\mathcal{O}({N}^3\mathcal{T})$ & \multirow{2}{*}{High computational complexity (LMMSE inversion)} \\ \cline{3-4}
 &  & DD-OAMP & $\mathcal{O}({N}^3\mathcal{T}+2N\mathcal{T}\log N)$ &  \\ \cline{3-5} 
 &  & MB-UAMP & $\mathcal{O}(L{N_s}^3+L{N_s}^2\mathcal{T})$ & \begin{tabular}[c]{@{}c@{}}Unitary preprocessing overhead and block-size\\ dependent trade-off\end{tabular} \\ \cline{3-5} 
 &  & DD-MAMP & $\mathcal{O}(N^2\mathcal{T})$ & Fails to fully exploit time-domain sparsity \\ \hline
\multirow{4}{*}{AFDM} & \multirow{4}{*}{Approximately sparse} & Gaussian AMP & $\mathcal{O}({N}^2\mathcal{T})$ & Performance degradation under high Doppler \\ \cline{3-5} 
 &  & MB-UAMP & \begin{tabular}[c]{@{}c@{}}$\mathcal{O}(L{N_s}^3+L{N_s}^2\mathcal{T})$\end{tabular} & \begin{tabular}[c]{@{}c@{}}High preprocessing overhead and sensitive to \\ block partitioning\end{tabular} \\ \cline{3-5} 
 &  & VV-EP & $\mathcal{O}(N^2L)$ & \begin{tabular}[c]{@{}c@{}} Iterative Cholesky optimization and sensitive to\\sparsity threshold tuning\end{tabular} \\ \cline{3-5} 
 &  & MAMP & $\mathcal{O}(N^2\mathcal{T})$ & Increased algorithmic complexity \\ \hline
IFDM & \begin{tabular}[c]{@{}c@{}}Dense and\\ right-unitarily invariant\end{tabular} & CD-MAMP & $\mathcal{O}(PN\mathcal{T}+2N\mathcal{T}\log N)$ & \begin{tabular}[c]{@{}c@{}}Universality limited to right-unitarily\\invariant channels\end{tabular} \\ \hline
IBST & \begin{tabular}[c]{@{}c@{}}Approximately dense \\ with sparse transform\end{tabular} & IBS-CD-MAMP & \begin{tabular}[c]{@{}c@{}}$\mathcal{O}(PN\mathcal{T}+2N\mathcal{T}\log N_s)$\end{tabular} & \begin{tabular}[c]{@{}c@{}}Performance–complexity trade-off governed by \\ block size\end{tabular} \\ \hline
RM & \begin{tabular}[c]{@{}c@{}}Dense\\ (universality class)\end{tabular} & CD-MAMP & \begin{tabular}[c]{@{}c@{}}$\mathcal{O}(PN\mathcal{T}+2N\mathcal{T}\log N)$\end{tabular} & Random transform generation and storage overhead \\ \hline
\end{tabular}
 \begin{tablenotes}
        \footnotesize
        \item[1] The main complexity orders indicate dominant asymptotic scaling for the overall detection process rather than exact computational cost, where $\mathcal{T}$ denotes the number of iterations, $P$ the number of channel paths, $L$ the number of blocks, and $N_s$ the dimension of each block.
      \end{tablenotes}
  \end{threeparttable}
}
\end{table*}

To address these challenges, orthogonal time-frequency space (OTFS) modulation has emerged as a promising candidate for reliable communication in high-mobility channels \cite{OTFS}. OTFS modulates symbols in the delay–Doppler (DD) domain, exploiting time-frequency diversity to mitigate multipath and Doppler effects \cite{OTFS2}. Corresponding iterative detectors have also been developed, such as cross-domain orthogonal AMP (CD-OAMP) \cite{CD-OAMP}, which iterates between the time and delay–Doppler domains, and delay–Doppler orthogonal AMP (DD-OAMP), which is designed in the delay–Doppler domain for OTFS to enable efficient and accurate detection \cite{DD-OAMP}. In addition, a multi-block unitary transformation–based approximate message passing (MB-UAMP) algorithm has been proposed for OTFS to reduce correlation among channel observations and improve convergence under severe Doppler conditions \cite{OTFS-UAMP}. However, OTFS suffers from limitations such as high modulation and detection complexity, as well as significant multi-user multiplexing overhead due to its two-dimensional (2D) modulation structure. Moreover, OAMP-based detectors for OTFS require high-complexity channel matrix inversions in the linear minimum mean square error (LMMSE) processing, which further limits their feasibility under hardware constraints. Although a delay-Doppler domain memory AMP (DD-MAMP) detector is proposed in \cite{DD-MAMP} as a substitute for the LMMSE in OAMP, it still fails to exploit the sparsity inherent in the time-domain channels. Likewise, MB-UAMP presents several implementation challenges. The computation and storage of unitary transformations, e.g., singular value decomposition (SVD), introduce considerable preprocessing overhead, and the selection of block size presents a trade-off among parallelization efficiency, memory usage, and residual correlation. 

In this context, affine frequency division multiplexing (AFDM) has been introduced as a novel multicarrier modulation scheme \cite{AFDM}. By applying an inverse discrete affine Fourier transform (IDAFT), AFDM warps data symbols into a chirp-modulated time–frequency domain, enabling full diversity in doubly selective channels. Compared to OTFS, AFDM offers a more hardware-friendly structure and lower implementation complexity, making it an attractive alternative to OFDM in high-mobility scenarios. Several detection methods have been developed. The Gaussian AMP detector in \cite{AFDM-GAMP} offers low-complexity detection for AFDM but suffers severe performance loss under fractional delay–Doppler channels due to strong correlation. The MB-UAMP previously applied to OTFS can be adapted to AFDM by using unitary preprocessing to reduce correlation and accelerate convergence \cite{AFDM-UAMP}, yet its preprocessing overhead and block-size trade-offs still pose implementation challenges. The vector-by-vector aided expectation propagation (VV-EP) detector for AFDM with index modulation achieves accurate detection with reduced computational load through sparse channel reconstruction \cite{VV-EP}, but it sacrifices mutual information and creates a measurable performance gap versus full-channel methods. Additionally, its iterative Cholesky optimization introduces additional overhead, and its accuracy is highly sensitive to threshold tuning across varying channel dispersion. Recently, a MAMP detector designed for AFDM under doubly dispersive channels has been proposed to better exploit channel statistics and improve convergence stability \cite{AFDM-MAMP}.

Although OTFS and AFDM represent notable advances, these schemes exhibit a fundamental limitation: they pair the multiplexing matrix with channel characteristics to enforce sparsity, creating a mismatch between actual channel statistics and the invariance assumptions required for optimal AMP-type detection algorithms, leading to performance degradation in practical correlated channels. To tackle this problem, interleave frequency division multiplexing (IFDM) has been proposed to produce a fully dense and right-unitarily invariant equivalent channel matrix via a random interleaver cascaded with inverse fast Fourier transform (IFFT), decoupling modulation from the channel to ensure each symbol experiences rich statistical fading. Combined with a cross-domain memory AMP detector (CD-MAMP) \cite{IFDM} that leverages super sparse time-domain channels and transform-domain right-unitary invariance, IFDM achieves near-capacity performance with significantly lower complexity than OAMP, outperforming OFDM, OTFS, and AFDM with their optimal detectors. Subsequently, an interleaved block-sparse transform (IBST) and its associated IBS-CD-MAMP detector were proposed to realize a sparse transform matrix \cite{IBST}, significantly reducing complexity and hardware requirements in resource-constrained scenarios. Building on this principle, random multiplexing (RM) has been proposed as a theoretically rigorous generalization \cite{RM}. Unlike prior schemes with deterministic structures or limited randomization, RM employs a channel-independent unitary random transform matrix that ensures the equivalent channel belongs to the universality class, which is strictly broader than right-unitary invariance. This guarantees replica-MAP optimality of AMP-type algorithms for arbitrary norm-bounded, spectrally convergent channels, relaxing restrictive channel assumptions and enabling reliable detection under realistic channel conditions \cite{RM}. For simplicity, we refer to the RM system employing the CD-MAMP detector as RM-MAMP. Table \ref{tab:M_D_Com} summarizes the representative modulation and detection techniques discussed above, and shows that RM-based modulation schemes better match the underlying assumptions of AMP-type algorithms than sparsity-based designs, thereby offering superior performance in high-mobility scenarios.

Despite these advantages, RM-MAMP still faces key challenges for practical deployment. First, large-scale random interleaving requires substantial storage and computational resources, making conventional implementations infeasible in hardware-constrained scenarios. Moreover, in ill-conditioned channels, particularly in small-scale settings, the effective diversity decreases, degrading detection accuracy. To characterize these challenging scenarios, we define {\it finite and severely ill-conditioned channels} as those with limited diversity and strong time–frequency correlation, which occur in small-scale, high-mobility, and severely time-varying scenarios. The ``finite" attribute refers to the limited system scale (specifically the interleaver sequence length), where performance loss induced by channel ill-conditioning grows more pronounced as this scale decreases. Here, the channel conditioning is quantified by the spectral condition number $\kappa(\bm H)=\sigma_{\max}(\bm H)/\sigma_{\min}(\bm H)$, expressed as $\kappa_{\mathrm{dB}}=20\log_{10}\kappa(\bm H)$, and we regard $\kappa_{\mathrm{dB}}\gtrsim80\ \mathrm{dB}$ as {\it finite and severely ill-conditioned}. Addressing the impact of such channels is therefore essential for bringing RM-MAMP into practical applications. Existing related works have verified the performance gain of interleaving and transform design for reliable signal detection \cite{IFDM,IBST,RM}. However, most of these works rely on conventional fully random interleaving schemes and focus on typical time-varying multipath channels, failing to simultaneously mitigate the storage overhead in hardware-constrained scenarios and robustness loss in {\it finite and severely ill-conditioned channels}.

To address the aforementioned challenges and bridge the gap between existing literature \cite{IFDM,IBST,RM}, this work makes the following major contributions:
\begin{itemize}[leftmargin=*]
  \item \textbf{Storage-efficient interleaver}: We propose two storage-efficient interleaver designs tailored for RM-MAMP. Specifically, we develop a Logistic chaotic mapping interleaver that generates random permutations from only two parameters, together with a quantitative parameter-selection criterion to ensure reliable randomness. To overcome the latency and precision-sensitivity limitations of chaotic interleavers, we further propose a dual-stage high-order permutation polynomial interleaver, which constructs deterministic random permutations using a small set of integer coefficients, achieving nearly identical BER as conventional designs with significantly reduced memory requirements and interleaver signaling overhead.
  \item \textbf{Highly reliable transform}: Built on the proposed storage-efficient interleaving architecture, we propose a highly reliable interleaved transform framework to mitigate performance loss in {\it finite and severely ill-conditioned channels}, which includes an interleaved phase perturbation transform and a multi-layer interleaved coupled transform. By introducing controlled phase randomness and multi-layer cross-domain coupling, the proposed highly reliable transform enhances the incoherence and diversity of the equivalent channel matrix, improving the reliability of iterative detection in {\it finite and severely ill-conditioned channels}.
\end{itemize}
The remainder of this paper is organized as follows. Section II introduces the system model and outlines the challenges. Section III presents the storage-efficient interleaving methods. Section IV describes the highly reliable transforms. Section V provides simulation results and performance analysis, and Section VI concludes the paper.

\begin{figure*}[htbp]
\centering
    \includegraphics[width=1\linewidth]{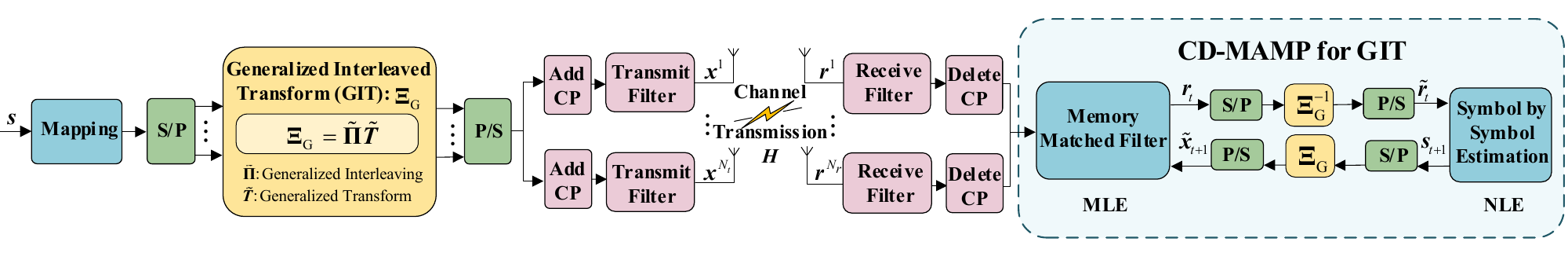}
    \vspace{-0.5cm}
\caption{A MIMO Generalized RM-MAMP communication system with $N_t$-antenna transmitter and $N_r$-antenna receiver.}
\label{model1}
\end{figure*}
\section{System Model and Challenges}
\subsection{Generalized RM-MAMP Communication System}
We first consider a multicarrier communication system, where the modulated signal sequence is denoted as ${\bm{x} = {\bm\Xi}{\bm s}}$. Here, ${\bm\Xi} \in {\mathbb{C}}^{N \times N}$ represents the multicarrier modulation matrix, and ${\bm{s}} \in {\mathbb{C}}^{N \times 1}$ is the symbol vector in the source domain. Then, a cyclic prefix (CP) is added to ${\bm{x}}$, and the system model is described as follows:
\begin{equation}
    \label{Eqn:onecon}
     {\bm{y}} = {\bm{H\Xi}}{\bm{s}} + \bm{n},
\end{equation}
where ${\bm{H}} \in {\mathbb{C}}^{N \times N}$ is the time-varying multipath channel matrix, and ${\bm{n}} \sim \mathcal{C}\mathcal{N}(0, \sigma^2 {\bm{I}}_N)$ is an additive white Gaussian noise (AWGN) vector. In this system, we assume that both ${\bm{y}}$ and ${\bm{H}}$ are known, while ${\bm{s}}$ is unknown. The goal is to design a multicarrier modulation matrix ${\bm\Xi}$ and a corresponding signal estimation algorithm to find a minimum mean square error (MMSE) estimate of the unknown vector ${\bm s}$. The system model can be expressed with the following two constraints:
\begin{equation}
    \label{Eqn:twocon}
    \Gamma: {\bm{y}} = {\bm{H\Xi}}{\bm{s}} + \bm{n}, \quad \Phi: s_i\sim P_s, \;\;\forall i.
\end{equation}
It is noted that the entries of $\bm{s}$ are independent and identically distributed (IID), i.e., $s_i\sim P_s$. As discussed in Section I, we select RM as ${\bm\Xi}$ in (\ref{Eqn:twocon}), as it outperforms other modulation schemes such as OFDM, OTFS, and AFDM. To estimate the signal ${\bm{s}}$, we employ CD-MAMP, which provides an MMSE estimate with both lower complexity and faster convergence compared to AMP and OAMP/VAMP \cite{IFDM,MAMP}. 

Now, we define a generalized RM-MAMP communication system. Fig.~\ref{model1} shows a multiple-input-multiple-output (MIMO) generalized RM-MAMP communication system, where S/P and P/S represent serial-to-parallel and parallel-to-serial conversion. The matrix ${\bm \Xi_{\mathrm G}}$ denotes the generalized interleaved transform (GIT), i.e., the multicarrier modulation process. Specifically, the GIT matrix ${\bm \Xi_{\mathrm G}}$ is defined as: ${\bm \Xi_{\mathrm G}}={\widetilde{\bm\Pi}}{\widetilde{\bm T}}$, where ${\widetilde{\bm\Pi}}$ denotes the generalized interleaving and ${\widetilde{\bm T}}$ represents the generalized transform. Then, we define the modulated signal as: $\bm{x}={\bm \Xi_{\mathrm G}}\bm{s}$. The CD-MAMP is applied for signal detection, effectively reducing the computational complexity of the iterative process by exploiting the structural properties of ${\bm H}$ and ${\bm \Xi_{\mathrm G}}$ \cite{MAMP}. In different types of interleaved transforms, the structure of the GIT matrix ${\bm \Xi_{\mathrm G}}$ varies, leading to distinct interpretations of the corresponding generalized interleaving ${\widetilde{\bm\Pi}}$ and generalized transform ${\widetilde{\bm T}}$. A detailed explanation for each case is provided below.
  \subsubsection{Original Interleaved Transform}
  In this scenario, we consider the original interleaved transform ${\bm \Xi_{\mathrm {ORI}}}$ from \cite{IFDM}. Specifically, ${\widetilde{\bm\Pi}}$ denotes the conventional interleaving operation—here represented by the random permutation matrix ${\hat{\bm{\Pi}}}$. Meanwhile, ${\widetilde{\bm T}}$ represents the standard signal transform, i.e., the $N$-point normalized IFFT matrix ${\bm F}^\mathrm H$. Then, the modulated signal $\bm{x}$ is defined as:
\begin{equation}
    {\bm x}={\bm \Xi_{\mathrm {ORI}}}{\bm s}={\hat{\bm{\Pi}}}{\bm F}^{\mathrm H}{\bm s}.
\end{equation}
This baseline configuration reflects both the randomness introduced by interleaving and the decorrelation achieved by the transform.
  \subsubsection{Interleaved Block-Sparse Transform}
  In this context, the product ${\widetilde{\bm\Pi}}{\widetilde{\bm T}}$ represents the interleaved block-sparse (IBS) transform, denoted as ${\bm \Xi_{\mathrm {IBS}}}$ \cite{IBST}. The IBS transform employs a novel interleaving strategy that integrates multiple low-dimensional matrices to form a sparse transform matrix, effectively reducing computational complexity and hardware demands, thereby enhancing efficiency in resource-limited scenarios. In this case, ${\widetilde{\bm\Pi}}$ and ${\widetilde{\bm T}}$ have different meanings compared to those of the baseline interleaved transform, and the signal $\bm{x}$ is defined as:
\begin{equation}
    {\bm x}={\bm \Xi_{\mathrm {IBS}}}{\bm s}={{\hat{\bm{\Pi}}}{\mathrm{diag}\{{\bm{{\Pi}}}_{1}{\bm{T}}_{1},...,{\bm{{\Pi}}}_{L}{\bm{T}}_{L}}\}}{\bm s},
\end{equation}
where ${\bm{\hat{\Pi}}}$ denotes the global interleaving matrix, and ${{\bm{\Pi}}_{l}}{\bm T}_{l}$ represents the $l$-th local interleaving transform matrix. Here, $L$ denotes the number of separated blocks. The transform matrix ${\bm T}_{l}$ represents either the inverse fast Fourier transform (IFFT) ${\bm F}^\mathrm H$ or the inverse fast Walsh–Hadamard transform (IFWHT) ${\bm W}^\mathrm H$. Note that the IFWHT replaces the complex-domain multiplications of the IFFT with real-domain additions and subtractions, thereby offering improved computational efficiency \cite{WHT}.
\subsection{Generalized RM-MAMP estimator}
For the generalized transform in RM, the generalized MAMP estimator is uniformly structured into four modules: a memory linear estimator (MLE), a non-linear estimator (NLE), a cross-domain transform (CDT), and the inverse CDT (ICDT):
\begin{subequations}\label{LE_NLE1}\begin{align}
    {\rm MLE\ }:&\;{\bm r}_t=\gamma_t({\bm X}_t), {\bm X}_t=[{\bm x}_1,...,{\bm x}_t];
    \\ {\rm ICDT}:&\;\tilde{\bm r}_t={\bm \Xi_{\mathrm G}}^{-1}{\bm r}_t,{\bm \Xi_{\mathrm G}}^{-1}={\bm \Xi}^{-1}_{\rm ORI}\;\text{or}\;{\bm \Xi}^{-1}_{\rm IBS};
    \\{\rm NLE}:&\;{\bm s}_{t+1}=\phi_t(\tilde{\bm r}_t);
    \\ {\rm CDT}:&\;\tilde{\bm x}_{t+1}={\bm \Xi_{\mathrm G}}{\bm s}_{t+1},{\bm \Xi_{\mathrm G}}={\bm \Xi}_{\rm ORI}\; \text{or}\; {\bm \Xi}_{\rm IBS}; 
\end{align}
\end{subequations}
where ${\bm \Xi}_{\rm ORI}$ and ${\bm \Xi}_{\rm IBS}$ denote the CDT for the original RM-MAMP and IBS-RM-MAMP, respectively. The output ${\bm r}_t$ is generated by the MLE $\gamma_t({\bm X}_t)$, which primarily depends on the channel matrix $\bm{H}$ and all previous estimations $[{\bm x}_1,...,{\bm x}_t]$. In the ICDT, the GIT matrix ${\bm \Xi_{\mathrm G}^{-1}}$ can be set to either matrix ${\bm \Xi}^{-1}_{\rm ORI}$ or matrix ${\bm \Xi}^{-1}_{\rm IBS}$, specifically:
\begin{equation}
   \! {\bm \Xi_{\mathrm G}}^{-1}=\left\{\begin{array}{l}{\bm \Xi}^{-1}_{\rm ORI}=\bm{F}{\hat{\bm{\Pi}}}^{-1},
    \\ 
    {\bm \Xi}^{-1}_{\rm IBS}=\mathrm{diag}\{{\bm{T}}_{1}^\mathrm{H}{\bm{{\Pi}}}_{1}^{-1},...,{\bm{T}}_{L}^\mathrm{H}{\bm{{\Pi}}}_{L}^{-1}\}{\hat{\bm{\Pi}}}^{-1}.
\end{array}\right.
\end{equation}
The estimate ${\bm s}_{t+1}$ in the NLE is produced by the non-linear symbol-by-symbol estimator $\phi_t(\tilde{\bm r}_t)$, where ${\bm s}_{t+1}$ is mapped to ${\bm x}_{t+1}$ by the GIT matrix ${\bm \Xi}_{\mathrm G}$. ${\bm \Xi_{\mathrm G}}$ employed in the CDT is defined as:
\begin{equation}
    {\bm \Xi_{\mathrm G}}=\left\{\begin{array}{l}{\bm \Xi}_{\rm ORI}={\hat{\bm{\Pi}}}\bm{F}^{\rm H},
    \\ 
    {\bm \Xi}_{\rm IBS}={{\hat{\bm{\Pi}}}{\mathrm{diag}\{{\bm{{\Pi}}}_{1}{\bm{T}}_{1},...,{\bm{{\Pi}}}_{L}{\bm{T}}_{L}}\}}.
\end{array}\right.
\end{equation}
\subsection{Challenges}
Despite the advantages of the RM-MAMP, several critical challenges must be addressed for its practical deployment:
\subsubsection{High Storage Costs in Practical Systems}
In large-scale communication systems, the interleaving operation in RM-MAMP imposes significant storage and computational demands. This process typically relies on generating pseudo-random sequences, which not only requires substantial storage but also increases processing requirements, particularly when dealing with long symbol sequences. This issue becomes more critical in hardware-constrained environments, where storage and computational resources are limited. Therefore, to mitigate the high memory overhead of interleaving in the RM-MAMP, it is crucial to develop storage-efficient solutions that minimize this overhead while maintaining system performance.
\subsubsection{Performance Degradation in Specific Channels}
Under severely time-varying channel conditions, particularly in small-scale scenarios, i.e., {\it finite and severely ill-conditioned channels} defined in Section I, the performance of the RM-MAMP system tends to degrade due to its sensitivity to rapid variations in the channel. Specifically, the estimation of the transmitted signal becomes increasingly inaccurate when the channel state changes rapidly, leading to significant performance loss. Such degradation becomes more evident in smaller systems, where limited resources amplify the effects of channel variability. Therefore, the design of a highly reliable RM-MAMP system capable of mitigating these performance losses under such challenging conditions is crucial.
\section{Storage-Efficient Interleaver Design}
To address the substantial storage burden and signaling overhead of the conventional interleaver in the RM-MAMP system, we introduce two storage-efficient designs that replace the need to store full interleaving sequences with the storage of only minimal mapping parameter sets, while preserving system performance.
\subsection{Logistic Chaotic Mapping-based (LCM) Interleaver}
To minimize the storage overhead of traditional interleavers that require storing full permutation sequences, one intuitive approach is to generate pseudo-random patterns dynamically from compact system parameters. Chaotic maps are attractive candidates for this purpose due to their inherent randomness and simplicity. The chaotic interleaver is derived from deterministic nonlinear systems, generating a highly decorrelated interleaving sequence with strong sensitivity to different initial conditions \cite{Chaos}. Compared to the conventional pseudo-random interleaver, which requires full-sequence storage, the chaotic interleaver achieves storage efficiency by constructing permutation patterns from compact system parameters, offering substantial memory savings. Among various chaotic maps, the Logistic map is particularly suitable due to its mathematical simplicity, strong randomness, and hardware-friendly structure, making it particularly well-suited for integration into the RM-MAMP system. Thus, we develop a Logistic chaotic mapping-based (LCM) interleaver tailored for the RM-MAMP framework, and the interleaving sequence is generated by iterating the Logistic map:
\begin{equation}
c_{n}=\psi_{c} c_{n-1} (1-c_{n-1}),c_0\in (0,1),\psi_{c} \in (3.5,4),
\end{equation}
where $c_0$ is a fixed initial value and $\psi_{c}$ is a parameter governing chaotic behavior. After generating a sequence of length $N$, the values $\{c_n\}_{n=0}^{N-1}$ are sorted in ascending order, and the indices of the sorted elements are recorded to form a permutation over the set of indexes $\{0, 1, ..., N-1\}$, which is then used as the interleaving pattern. The key advantage of this approach lies in its minimal storage requirement: only two values $(\psi_c,c_0)$ need to be stored and transmitted, in contrast to the traditional interleaver that requires a full sequence of $N$ indices. 

Building on this principle, we propose a parameter-selection criterion for the LCM interleaver employed in the RM-MAMP system. Rather than choosing chaotic parameters arbitrarily, we evaluate candidate parameter pairs $(\psi_c,c_0)$ by a combined criterion that uses a permutation displacement metric $\mathcal{D}_s$ and a discrete entropy measure $\mathcal{H} _s$. Let $\bm{s}=[s_0,\ldots,s_{N-1}]$ denote the interleaving index sequence obtained by ranking the chaotic samples \(\{c_n\}_{n=0}^{N-1}\). Then $\mathcal{D}_s$ is defined to quantify index displacement: 
\begin{equation}
\mathcal{D}_s=\frac{1}{N}\sum_{i=0}^{N-1}|s_i-i|. 
\end{equation}
To quantify the finite-precision unpredictability of the chaotic samples, we partition the unit interval into $B$ bins, count the empirical frequencies $n_b$ into each bin, and set the empirical probabilities $p_b=n_b/N$. The normalized discrete entropy is: 
\begin{equation}
\mathcal{H} _s = -\frac{1}{\log B}\sum_{b=1}^{B} p_b\log p_b, \quad\mathcal{H} _s\in[0,1], 
\end{equation}
where $\mathcal{H} _s$ closer to 1 indicates a near-uniform chaotic sample distribution. In practice, we set $B$ according to the sample length and implementation precision (e.g., \(B\approx\sqrt{N}\) to ensure stable entropy estimation). We then combine the two normalized metrics into a single composite score:
\begin{equation}
S(\psi_c,c_0) \;=\; w_\mathcal{D}\mathcal{D}_s+w_\mathcal{H}\mathcal{H} _s,
\end{equation}
where \(w_\mathcal{D},w_\mathcal{H}\ge 0\) and \(w_\mathcal{D}+w_\mathcal{H}=1\). Candidate parameters are accepted if they maximize $S$ over the search set and satisfy a short-period constraint, i.e., no repetition occurs within a pre-specified period \(\mathcal{T}_{\min}\). In our design, the Logistic map operating in the fully chaotic region with $\psi_{c} \in (3.5,4)$ inherently exhibits a positive Lyapunov exponent, strong ergodicity, and near-uniform distribution \cite{Chaos_Tutorial,Chaos_Dynamical}, which serves as the core source of randomness for the LCM interleaver. The proposed dual-metric composite criterion is introduced as a robustness constraint to suppress potential structured residuals in finite-precision implementation. The equal weight setting \(w_\mathcal{D}=w_\mathcal{H}=0.5\) is adopted to balance the global dispersion of the permutation sequence and the uniformity of the chaotic sample distribution, avoiding extreme cases under single-metric constraints. Preliminary simulation results confirm negligible performance differences across weight combinations. This criterion thus delivers a quantitatively controlled design, with the LCM interleaver serving as a storage-extreme baseline for the RM-MAMP system.

Despite the storage advantage noted above, the LCM interleaver exhibits some practical limitations that undermine its suitability for high-throughput implementations.
\begin{itemize}[leftmargin=*]
  \item {Interleaving complexity}: Generating an $N$-length permutation requires ${\cal O}(N)$ mapping iterations followed by ${\cal O}(N\log N)$ global ranking. The complexity grows superlinearly with the frame length, and fixed-point calculation for chaotic randomness amplifies hardware overhead.
  \item {Sorting latency}: The ranking operation necessitates a comparator-tree whose pipeline depth scales approximately as $\log _{2}N$, thus end-to-end delay contains a non-negligible ${\cal O}(\log N)$ component in addition to the iteration pipeline.
  \item {Parallel adaptability}: Sample generation is sequential, and global ranking in the LCM interleaver requires full-data coordination, which limits fine-grained parallelization and reduces achievable throughput on parallel hardware.
  \item {Finite-precision sensitivity}: Chaotic orbits suffer from severe randomness loss under low-bit-width fixed-point implementation, which undermines hardware efficiency and introduces non-negligible implementation risk.
\end{itemize}
Consequently, the memory savings of the LCM interleaver are offset by elevated computational cost, non-negligible latency, poor parallel scalability, and critical finite-precision implementation risks, making it suboptimal for reliable low-latency or highly parallel hardware deployments.
\subsection{Dual-Stage High-Order Permutation Polynomial-based (DPP) Interleaver}
To address the practical limitations of the LCM interleaver, we propose a dual-stage high-order permutation polynomial-based (DPP) interleaver that realizes low-latency, storage-efficient permutations using only a compact set of integer coefficients. Let the base interleaving function $\pi(\bm s)$ be defined as an $m$-th order polynomial over the integer ring $\mathbb{Z}_{\mathit{N}}$, where $\mathit{N}$ is the sequence length. Then, we define $\pi(\bm s)$ as:
\begin{equation}
    \label{Eqn:HPPs} \pi(\bm s)=\displaystyle\sum_{i=0}^{m}a_is^i\pmod N,s\in \mathbb{Z}_{\mathit{N}}.
\end{equation}

To ensure $\pi(s)$ is a complete permutation (bijective mapping) over $\mathbb{Z}_N$ \cite{HPP}, our high-order permutation polynomial (HPP) coefficient design is rooted in the \textit{Chinese Remainder Theorem} (CRT) and classic permutation polynomial theory \cite{Finite_Fields,QPP}. By the CRT, $\pi(s)$ permutes $\mathbb{Z}_N$ if and only if it is a permutation over every prime power factor of $N$ \cite{Rosen2010Elementary}, guiding our design as follows:
\begin{itemize}[leftmargin=*]
  \item Linear term invertibility: The linear coefficient $a_{1}$ must be coprime with $N$, which is essential to guarantee the uniqueness in the mapping: $\gcd(a_{1}, N) = 1$. This ensures $a_1$ is invertible modulo all prime factors of $N$, satisfying the core permutation requirement via the CRT \cite{QPP}.
  \item Higher-order term divisibility: Let $\mathcal{P}_{\text{odd}} = \{p_{i} \in \mathbb{P} : p_{i} \mid N, p_{i} > 2\}$ denote the set of odd prime divisors of $N$. Then, all higher-order coefficients must satisfy the constraint:
\begin{equation}
a_{i} \equiv 0 \pmod{p}, \forall i \in \{2, 3, \ldots, m\}, p \in \mathcal{P}_{\text{odd}}.
\end{equation}
This reduces the polynomial to a linear form modulo each odd prime factor of $N$. Combined with the linear term, this form is a permutation over odd prime fields \cite{Finite_Fields}, ensuring full permutation over all odd prime power factors of $N$.
  \item Parity consistency: If $N = 2^q \cdot N_1$ ($q \geq 2$, $N_1$ odd), two conditions must hold to prevent collisions when reducing the polynomial modulo 2 or 4. First, the difference between the sum of odd-degree coefficients and even-degree coefficients (excluding $a_0$) is odd:
  \begin{equation}
  \label{Eqn:M2}
  \sum_{i=0}^{\lfloor (m-1)/2 \rfloor} a_{2i+1} - \sum_{i=1}^{\lfloor m/2 \rfloor} a_{2i} \equiv 1 \pmod{2}.
  \end{equation}
  Secondly, the sum of coefficients with degree $\geq 2$ is even:
  \begin{equation}
  \label{Eqn:M4}
  \sum_{i=2}^{m} a_i \equiv 0 \pmod{2}.
  \end{equation}
These conditions guarantee permutation over $\mathbb{Z}_{2^q}$ for any $q \geq 2$ \cite{Mullen2013Handbook}, completing the CRT requirements for full bijectivity over $\mathbb{Z}_N$.
\end{itemize}

For illustration, we consider coefficient selection for $N=256$. Taking $N=256=2^8$ (with no odd prime divisors) and a 4th-order polynomial $\pi(s)=a_4s^4+a_2s^2+a_1s+a_0$ as an example: (1) Choose $a_1=5$ (ensuring $\gcd(5,256)=1$); (2) The higher-order term divisibility is automatically satisfied since there are no odd prime divisors; (3) For parity consistency: choose $a_4=1$ and $a_2=3$, where their sum $1+3=4$ is even to satisfy equation (\ref{Eqn:M4}), and verify equation (\ref{Eqn:M2}): $5 - (1+3) = 1 \equiv 1 \pmod{2}$; (4) Set $a_0=0$, as this term is arbitrary and has no impact on bijectivity. Then the coefficients are $a_4=1, a_2=3, a_1=5, a_0=0$. Note that this is only one valid set of coefficients, and all other coefficient values satisfying the above constraints are equally applicable.

Then, two such base HPPs are employed in cascade, i.e., a dual-stage HPP (DPP) interleaver, combined with structured reshaping operations:
\begin{equation}
\bm{\hat s} = (\mathcal{R}_2 \circ \pi^{(2)} \circ \mathcal{R}_1 \circ \pi^{(1)})(\bm{s}).
\end{equation}
Here, let $k$ denote the $k$-th stage of the HPP in the Dual-HPP interleaver. The corresponding permutation polynomial at stage $k$ is given by: 
\begin{equation}
\pi^{(k)}(\bm{s}) = \sum_{i=0}^{m} a_i^{(k)} s^i \pmod N, k=1,2.
\end{equation} 
In addition, we introduce the operator $\mathcal{R}_k = \mathcal{R}_k(\cdot; L_k, C_k)$ to reshape the input sequence into an $L_k \times C_k$ matrix by writing row-wise and reading column-wise.
Moreover, we define the operator $\circ$ to denote the function composition over the matrix space within the Dual-HPP interleaver framework, specifically: 
$(\mathcal{M}\circ\mathcal{N})(\bm{s})=\mathcal{M}(\mathcal{N}(\bm{s})).$
This construction achieves excellent sequence randomness and decorrelation through the polynomial-domain mapping, while significantly reducing the high storage requirement, as only two small sets of coefficients $\{a_i^{(1)}\}_{i=0}^{m}$ and $\{a_i^{(2)}\}_{i=0}^{m}$ ($m\ll N$) need to be stored. Compared with the LCM interleaver discussed previously, the proposed DPP interleaver achieves a more hardware-efficient operation.

\begin{table*}[htbp]
\centering
\caption{Comparison of Original Random, LCM, and DPP Interleavers (Frame Length $= 2048$).}
\label{tab:LCM_DPP_Com}
\scalebox{0.76}{
\begin{threeparttable} 
\begin{tabular}{|c|c|c|c|}
\hline
Metric & Original Random Interleaver & LCM Interleaver & DPP Interleaver \\ \hline
\multicolumn{4}{|c|}{{Core Storage Efficiency}} \\ \hline
Algorithmic Storage Complexity & $\mathcal{O}(N)$ & $\mathcal{O}(1)$ & $\mathcal{O}(1)$ \\ \hline
Practical On-Chip Storage & 2.8 KB (full permutation table) & 8 B (2 fixed parameters) & 16 B (2 sets of coefficients) \\ \hline
Transceiver Signaling Overhead & Full $N$-length sequence & 2 parameters & Small coefficient sets\\ \hline
\multicolumn{4}{|c|}{{Hardware Implementation Overhead}} \\ \hline
Random Memory Access per Frame & 4096 times (severe bottleneck) & 0 & 0 \\ \hline
Sequential Arithmetic Operations & ~0 & ~24576 operations (32-bit fixed-point) & ~6144 operations \\ \hline
Total Equivalent Overhead\tnote{1} & $\approx$73728 equivalent operations & $\approx$24576 equivalent operations & $\approx$6144 equivalent operations \\ \hline
Memory Access Pattern & Fully random (high bus overhead) & Fully sequential & Streamable (no frame buffer) \\ \hline
Fixed-Point Bit-Width Requirement & No precision constraint & $\geqslant$32-bit (to avoid precision loss) & No precision constraint \\ \hline
\multicolumn{4}{|c|}{{Latency, Parallelism and Stability}} \\ \hline
Pipelined End-to-End Latency & ~8192 cycles & 11 cycles + precision overhead & 1 cycle \\ \hline
Parallel Adaptability & Poor & Poor & High \\ \hline
Finite-Precision Randomness Loss & None & Severe under fixed-point & None \\ \hline
\end{tabular}
\begin{tablenotes}
\footnotesize
\item[1] Random memory access is the dominant hardware bottleneck, with an equivalent overhead 15 to 20 times higher than sequential arithmetic operations.
\end{tablenotes}
\end{threeparttable}
}
\end{table*}
Table \ref{tab:LCM_DPP_Com} quantifies the implementation performance of the proposed LCM and DPP interleavers against the original random interleaver adopted in the baseline RM-MAMP system. Both proposed storage-efficient designs eliminate the dominant hardware bottleneck of random memory access inherent to full permutation table lookup, reduce on-chip storage and transceiver signaling overhead from the $\mathcal{O}(N)$ scale of the baseline to $\mathcal{O}(1)$, and preserve the permutation randomness required to maintain the universality and detection optimality of RM-MAMP system. The LCM interleaver achieves extreme storage compression with only two fixed parameters, but suffers from severe finite-precision sensitivity in practical hardware deployment, with its total arithmetic operations comprising $N$ chaotic mapping iterations and $N{\log_2}N$ sorting comparisons. This requires high-bit-width calculation and additional cycle compensation logic to avoid chaotic orbit degradation and randomness loss, which increases actual hardware overhead and undermines hardware efficiency. The frame-level sorting operation of the LCM interleaver also restricts its parallelization capability and low-latency performance. In contrast, the DPP interleaver is insensitive to finite precision. It removes the sorting step via deterministic algebraic mapping, delivering fixed single-cycle latency, minimal hardware overhead, and full parallelization, making it ideal for storage-efficient random multiplexing.

\subsection{Universality Compliance of LCM and DPP Interleavers}
The replica-MAP optimality of RM-MAMP relies on the equivalent channel matrix belonging to the universality class \(\mathscr{U}\) defined in RM \cite{RM}. This section verifies that the proposed LCM and DPP interleavers satisfy the core criteria for membership in \(\mathscr{U}\), while retaining their low-storage advantages.
\subsubsection{Core Criteria for Universality Class \(\mathscr{U}\) in RM}
The original definition of \(\mathscr{U}\) for RM in \cite{RM} can be summarized by two core properties: (1) Spectral Convergence: The equivalent channel matrix \(\boldsymbol{H}_{\mathrm{eff}} = \boldsymbol{H}\boldsymbol{\Xi}\) has a bounded spectral norm \(\|\boldsymbol{H}_{\mathrm{eff}}\|_2 \lesssim 1\) \cite{RM}. In this paper, this property is verified by the convergence of the empirical spectral distribution (ESD) of \(\boldsymbol{H}_{\mathrm{eff}}^{\mathrm{H}}\boldsymbol{H}_{\mathrm{eff}}\) to the Marchenko-Pastur (M-P) distribution, which confirms that the large-dimensional \(\boldsymbol{H}_{\mathrm{eff}}\) has the universal statistical properties required for valid state evolution (SE) analysis. (2) Input Isotropy Structure: \(\boldsymbol{H}_{\mathrm{eff}}\) satisfies the permutation-invariant structure \(\boldsymbol{H}_{\mathrm{eff}} = \boldsymbol{J}\boldsymbol{D}\) \cite{RM}. Here, \(\boldsymbol{D}\) is a diagonal matrix with IID uniform random phases over \([0,2\pi)\), \(\boldsymbol{J}\) is a deterministic spectrally convergent matrix with bounded norm, and entries of \(\boldsymbol{J}^{\mathrm{H}}\boldsymbol{J}\) are asymptotically uniform as \(N \to \infty\). This ensures the asymptotic IID Gaussianity of the error terms in CD-MAMP, which is a core prerequisite for accurate SE analysis.

The membership of \(\mathscr{U}\) is primarily determined by these two criteria, which guide our subsequent analysis. Accordingly, LCM and DPP interleavers should generate permutations asymptotically statistically equivalent to uniform random permutations, thus preserving membership in \(\mathscr{U}\).
\subsubsection{Universality Compliance of the LCM Interleaver}
The LCM interleaver meets both core properties of \(\mathscr{U}\) in RM via the intrinsic dynamics of the fully chaotic Logistic map and our parameter optimization criterion: (1) Spectral Convergence Guarantee: In the fully chaotic regime with \(\psi_c \in (3.5,4)\), the Logistic sequence exhibits near-zero autocorrelation at non-zero lags, generating an aperiodic permutation matrix with asymptotically orthogonal columns \cite{Chaos_Sequence_Overview}. This ensures the ESD of \(\boldsymbol{H}_{\mathrm{eff}}^{\mathrm{H}}\boldsymbol{H}_{\mathrm{eff}}\) converges to the M-P distribution in the large-system limit. (2) Input Isotropy Structure Guarantee: The fully chaotic Logistic map satisfies the Birkhoff ergodic theorem, with samples asymptotically uniform over \((0,1)\) \cite{Chaos_Tutorial,Chaos_Dynamical}. The permutation sequence converges to the uniform Haar measure on the symmetric group \(S_N\) as \(N \to \infty\), statistically equivalent to uniform random permutation \cite{Chaos_Dynamical}, thus satisfying the permutation-invariant structure requirement of \(\mathscr{U}\).
\subsubsection{Universality Compliance of the DPP Interleaver}
The DPP interleaver satisfies both core properties via the mathematical properties of the dual-stage bijective permutation polynomial: (1) Spectral Convergence Guarantee: The dual-stage cascaded structure eliminates inherent structural correlation in single-stage permutation polynomials, yielding an aperiodic permutation matrix with extremely low differential uniformity. For any fixed \(N\), the inner product expectation between distinct columns is 0, with variance decaying at a rate of \(1/N\) \cite{Permutation_Poly_Turbo_Bounds}, ensuring the ESD of \(\boldsymbol{H}_{\mathrm{eff}}^{\mathrm{H}}\boldsymbol{H}_{\mathrm{eff}}\) converges to the M-P distribution. (2) Input Isotropy Structure Guarantee: The dual-stage permutation polynomial satisfies bijective constraints over \(\mathbb{Z}_N\), forming a measure-preserving bijection with full ergodicity. As \(N \to \infty\), the permutation sequence converges to the uniform Haar measure on \(S_N\), statistically identical to a uniform random permutation \cite{Finite_Fields}, meeting the permutation-invariant structure requirement of \(\mathscr{U}\).
\begin{figure}[htbp]
\centering
 \includegraphics[width=1.0\linewidth]{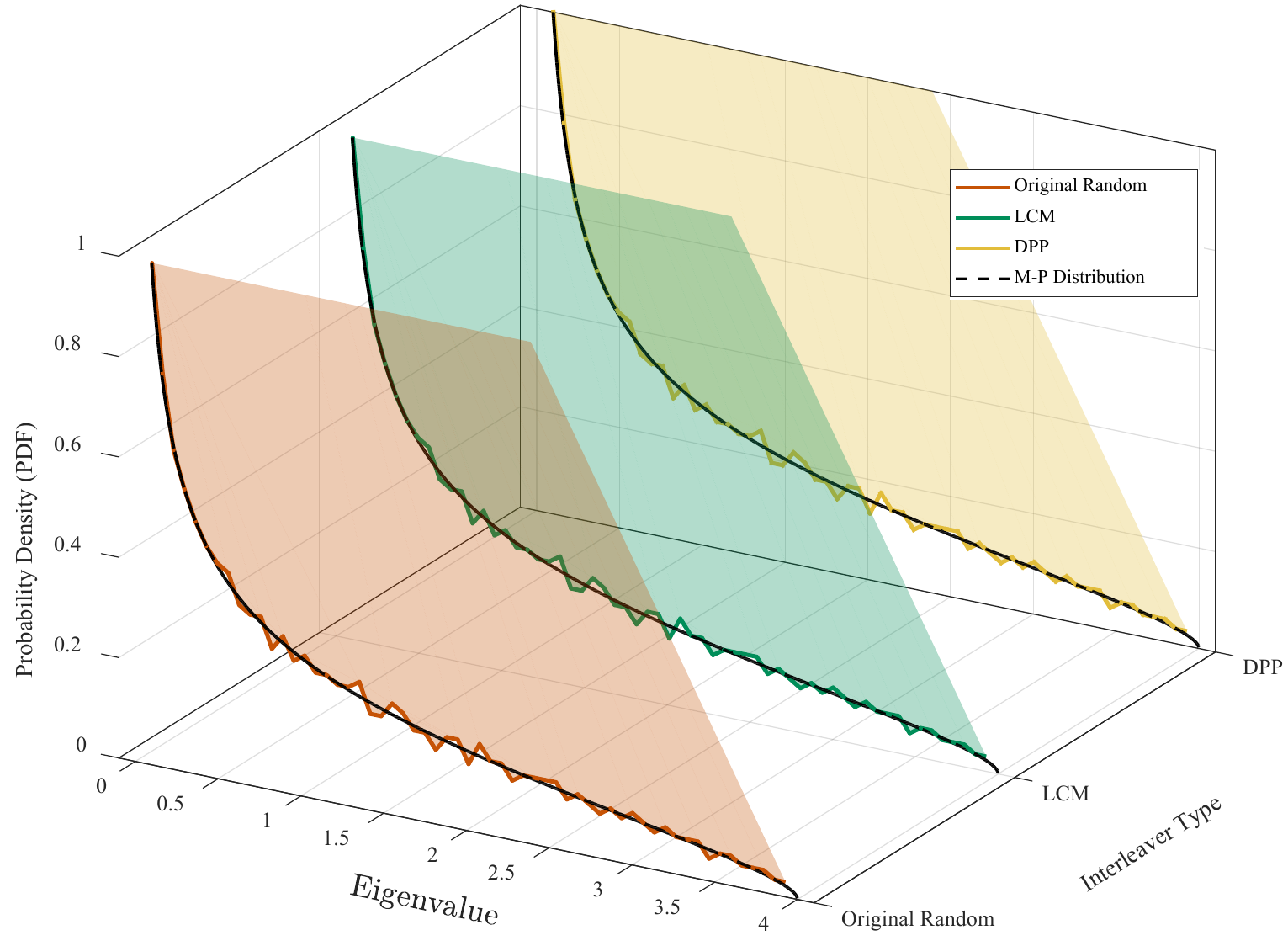}
\caption{Empirical spectral distribution (ESD) comparison of the original random, LCM, and DPP interleavers with 1024 subcarriers.}
\label{MR1}
\end{figure}

In summary, both the LCM and DPP interleavers satisfy the two core properties for the universality class \(\mathscr{U}\). As shown in Fig.~\ref{MR1}, the empirical spectral distribution (ESD) of the Gram matrices \(\boldsymbol{H}_{\mathrm{eff}}^{\mathrm{H}}\boldsymbol{H}_{\mathrm{eff}}\) for the original random, LCM, and DPP interleavers all closely match the Marchenko-Pastur (M-P) distribution, with discrepancies below $10^{-8}$ across the three interleavers. This directly validates the spectral convergence and boundedness property of \(\mathscr{U}\) in LCM and DPP interleavers and corroborates the satisfaction of the input isotropy structure requirement.

\section{Highly Reliable Transform Design}
To address the performance degradation of the RM-MAMP system in {\it finite and severely ill-conditioned channels} defined in Section I, we introduce two high-reliability transforms to reconstruct the modulation and demodulation process in RM-MAMP, i.e., the interleaved phase perturbation transform and the interleaved multi-layer coupled transform. These transforms aim to increase the randomness and incoherence of the effective channel matrix, thereby improving the reliability of signal recovery in the RM-MAMP system under challenging communication scenarios.
\subsection{Interleaved Phase Perturbation Transform (IPPT)}
The IPPT employs random phase perturbation to further disrupt the residual structure in the equivalent channel matrix, thereby achieving diversity gains in severely multipath-fading channels. Let $N$ denote the length of the information sequence, then we define a diagonal phase perturbation matrix ${\bm\Phi}^{x}\in \mathbb{C}^{N\times N}$ as:
\begin{equation}
\bm{\Phi}^{x}
 = \mathrm{diag}\left(e^{-j\theta_n},n = 0, \dots, N-1\right),
\end{equation}
where $x\in{{\mathrm{T}},{\mathrm{S}}}$, i.e., \(\bm{\Phi}^\mathrm{T}\) and \(\bm{\Phi}^\mathrm{S}\), correspond to transform-domain and source-domain phase perturbations, respectively. $\theta_n \sim \mathcal{U}(0, 2\pi)$ denotes a random variable uniformly distributed over $[0, 2\pi)$. Based on the shared structure of the original interleaved transform and the interleaved block-sparse transform, we consider phase perturbation applied to the composite mapping $\hat{\bm{\Pi}} \bm{T}$, where $\hat{\bm{\Pi}}$ is a random permutation matrix and $\bm{T}$ denotes either ${\bm F}^\mathrm H$ or ${\bm W}^\mathrm H$ (see Section II for definitions). Specifically, we examine three placements of the phase perturbation.
\subsubsection{Dual-Side Phase Perturbation}
We first introduce phase randomness to both the interleaving and modulation stages, which provides the most comprehensive effect, i.e., IPPT-DUAL: 
\begin{equation}
\bm{\Xi}^{\mathrm{D}}_\mathrm{IPP}=\bm{\Phi}^\mathrm{T}\hat{\bm{\Pi}} \bm{T}\bm{\Phi}^\mathrm{S},
\end{equation}
where $\bm{\Phi}^\mathrm{T}, \bm{\Phi}^\mathrm{S}\in \mathbb{C}^{N\times N}$ are diagonal matrices with $\theta_n \sim \mathcal{U}(0, 2\pi)$ as phase shifts, respectively, in the transform and source domains. This achieves comprehensive diversity gains by eliminating residual deterministic structures across both time and frequency domains.

\subsubsection{Inner-Side Phase Perturbation}
We then introduce the phase perturbation to the modulation stage. For an input symbol vector $\bm{s}$, the operations follow the right-to-left order implied by the matrix product:
\begin{equation}
\bm{\Xi}^{\mathrm{S}}_\mathrm{IPP}=\hat{\bm{\Pi}} \bm{T}\bm{\Phi}^\mathrm{S},
\end{equation}
where $ \bm{\Phi}^\mathrm{S}\in \mathbb{C}^{N\times N}$ is a diagonal matrix as phase shifts in the source domain, which affects signals before the multicarrier modulation $\hat{\bm{\Pi}} \bm{T}$. For simplicity, we refer to this scheme as IPPT-IN. In practice, the effectiveness of IPPT-IN depends on the pre-modulation phase statistics. Specifically, IPPT-IN is effective when source-domain symbol phases are concentrated or structured, such as in short frames, weakly interleaved blocks, repeated pilot sequences, or symbol mappings that yield limited phase diversity. In these cases, per-subcarrier phase rotations increase phase diversity and reduce residual frequency-domain correlation. Conversely, IPPT-IN yields negligible benefit when the source-domain phases are already widely dispersed (e.g., after strong scrambling or for mapping that produces diverse subcarrier phases) or when the symbol alphabet is phase-limited (e.g., BPSK or QPSK). Under such conditions, inner-side rotations do not materially change the phase statistics and therefore provide little additional decorrelation. In this work, because the considered symbol distribution and interleaving produce sufficiently dispersed source-domain phases, we find IPPT-IN to be ineffective and thus focus on outer-side perturbation.

\subsubsection{Outer-Side Phase Perturbation}
We employ the phase perturbation after the interleaving stage: 
\begin{equation}
\bm{\Xi}^{\mathrm{T}}_\mathrm{IPP}=\bm{\Phi}^\mathrm{T}\hat{\bm{\Pi}} \bm{T},
\end{equation}
where $ \bm{\Phi}^\mathrm{T}\in \mathbb{C}^{N\times N}$ is applied in the transform domain to mitigate residual inter-symbol interference patterns, which is considered to be IPPT-OUT for brevity. By multiplying the transmitted samples with a diagonal phase matrix, the relative phases of different channel columns are randomized, which reduces column coherence and makes the equivalent matrix statistics closer to the right-unitary invariance assumed by MAMP. Thus, IPPT-OUT increases effective diversity and improves the stability and convergence of RM-MAMP in {\it finite and severely ill-conditioned channels} defined in Section I, with only modest implementation overhead, making it a preferable choice for adoption in this work.

\subsection{Interleaved Multi-Layer Coupled Transform (IMCT)}
Beyond random phase perturbation, we further enhance reliability through a transform domain coupling strategy that spreads the energy of each transmitted symbol across multiple subcarriers and time samples. This idea can be traced back to the energy-spreading transform (EST) \cite{IEST0,IEST1}, which redistributes transmitted energy over multiple subcarriers to improve frequency diversity and has since been widely adopted in various modulation schemes. 

Building on this principle, we propose the interleaved multi-layer coupled transform (IMCT), which generalizes single-stage spreading into a multi-layer cascaded structure of $K$ coupled transform layers combining interleaving, alternating cross-domain unitary transforms, and optional outer-phase rotations:
\begin{equation}
    \begin{aligned}
\!\bm{\Xi}_\mathrm{IMC}=&\prod_{k=1}^{K}\bigl(\boldsymbol{\Phi}_{k}\,\hat{\boldsymbol{\Pi}}_{k}\,{\boldsymbol T}^{\eta_k}_{k}\bigr)\\
=&\boldsymbol{\Phi}_{K}\,\hat{\boldsymbol{\Pi}}_{K}\,{\boldsymbol T}^{\eta_K}_{K}\dots\boldsymbol{\Phi}_{2}\,\hat{\boldsymbol{\Pi}}_{2}\,{\boldsymbol T}^{(\eta_2)}_{2}\boldsymbol{\Phi}_{1}\,\hat{\boldsymbol{\Pi}}_{1}\,{\boldsymbol T}^{\eta_1}_{1},
\end{aligned}
\end{equation}
where ${\boldsymbol T}^{\eta_k}_{k}$ alternates between the transform matrix (e.g., ${\bm F}$ or ${\bm W}$ in Section II) and its Hermitian inverse (e.g., ${\bm F}^\mathrm H$ or ${\bm W}^\mathrm H$) across layers. Equivalently, if \({\bm T}_k^{\eta_k}\) is the forward transform at an odd layer \(k\), then the corresponding matrix at the even layer is its Hermitian transpose, i.e., \({\bm T}_{k+1}^{\eta_{k+1}}=\bigl({\bm T}_k^{\eta_k}\bigr)^{\mathrm H}\).
 $\hat{\boldsymbol{\Pi}}_{k}$ denotes the layer-specific permutation, and $\boldsymbol{\Phi}_{k}$ is an optional diagonal phase perturbation matrix applied in the source domain or transform domain, as discussed in IPPT-OUT (Section IV-1). In this framework, $\boldsymbol{\Phi}_{k}$ is not always required, i.e., if the interleaved multi-layer coupling itself sufficiently compensates for the performance degradation, $\boldsymbol{\Phi}_{k}$ can be replaced by an identity matrix ${\bm I}$. Moreover, when $K=1$, the IMCT reduces to the outer-side phase perturbation structure IPPT-OUT, i.e., $\bm{\Xi}^{\mathrm{T}}_\mathrm{IPP}$. 
Therefore, we regard IPPT-OUT and IMCT as specific instances of a unified, highly reliable (HR) transform framework. Specifically, we define \(\boldsymbol{\Xi}_{\mathrm{HR}}\) as the HR transform:
\begin{equation}
\label{eq:HR_transform_aligned}
\begin{aligned}
\!\boldsymbol{\Xi}_{\mathrm{HR}}
&=\prod_{k=1}^{K}\Bigl(\boldsymbol{\Phi}_{k}\,\hat{\boldsymbol{\Pi}}_{k}\,{\boldsymbol T}_{k}^{\eta_k}\Bigr),
\quad {\boldsymbol T}_{k}^{\eta_k}\in\{{\boldsymbol T}_k,\,{\boldsymbol T}_k^{\mathrm H}\} \\
&= 
\begin{cases}
\displaystyle \boldsymbol{\Phi}_{1}\,\hat{\boldsymbol{\Pi}}_{1}\,{\boldsymbol T}_{1}^{\eta_1}, & K=1(\text{IPPT-OUT}) ,\\
\displaystyle \boldsymbol{\Phi}_{K}\,\hat{\boldsymbol{\Pi}}_{K}\,{\boldsymbol T}^{\eta_K}_{K}\dots\boldsymbol{\Phi}_{1}\,\hat{\boldsymbol{\Pi}}_{1}\,{\boldsymbol T}^{\eta_1}_{1}, & K>1(\text{IMCT}).
\end{cases}
\end{aligned}
\end{equation}

\begin{figure}[htbp]
\centering
 \includegraphics[width=1.0\linewidth]{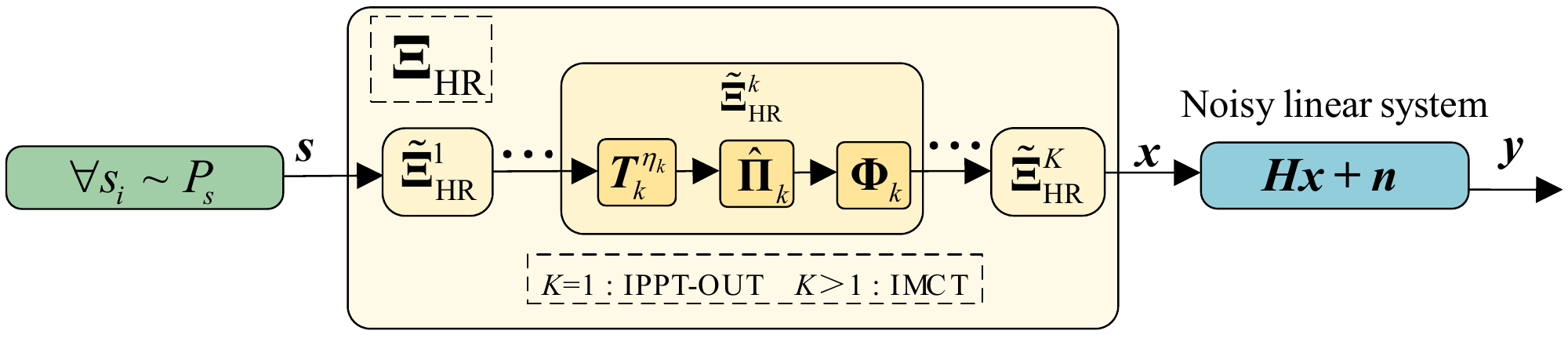}
\caption{A transmitter-side representation of the highly reliable multicarrier modulation linear system.}
\label{model2}
\end{figure}

In this HR transform, when \(K=1\), \(\boldsymbol{\Xi}_{\mathrm{HR}}\) reduces to the single-layer interleaved transform (i.e., IPPT-OUT); when \(K>1\), \(\boldsymbol{\Xi}_{\mathrm{HR}}\) corresponds to the multi-layer coupled interleaved transform (i.e., IMCT). The HR structure disperses the energy of each transmitted symbol across additional approximately independent channel observations, reduces column coherence of the equivalent channel matrix, and preserves the right-unitary invariance assumed by RM-MAMP. With complexity growing approximately linearly with $K$ and negligible storage overhead when storage-efficient interleavers in Section III are used, the HR transform provides a practical framework to improve detection robustness in {\it finite and severely ill-conditioned channels}. Subsequently, within the generalized RM-MAMP communication system in Section II, the GIT matrix ${\bm \Xi_{\mathrm G}}$ is defined as the highly reliable (HR) transform under the HR design for enhanced RM-MAMP. The transmit-side representation of the HR multicarrier modulation linear system is redefined as follows:
\begin{equation}
 \bm{x} \;=\; {\bm \Xi}_{\mathrm{HR}}\,\bm{s}=\prod_{k=1}^{K}\Bigl(\boldsymbol{\Phi}_{k}\,\hat{\boldsymbol{\Pi}}_{k}\,{\boldsymbol T}_{k}^{\eta_k}\Bigr)\bm{s},
\end{equation}
where the GIT matrix ${\bm \Xi_{\mathrm G}}$ in the generalized RM-MAMP is denoted as: ${\bm \Xi_{\mathrm G}}={\bm \Xi}_{\mathrm{HR}}$ , which is used to perform the HR transmission. Fig.~\ref{model2} illustrates the highly reliable multicarrier modulation linear system at the transmitter side, where $\widetilde{\bm \Xi}_{\mathrm{HR}}^{k}$ denotes the $k$-th coupled transform layer.

\begin{figure}[htbp]
\centering
 \includegraphics[width=1.0\linewidth]{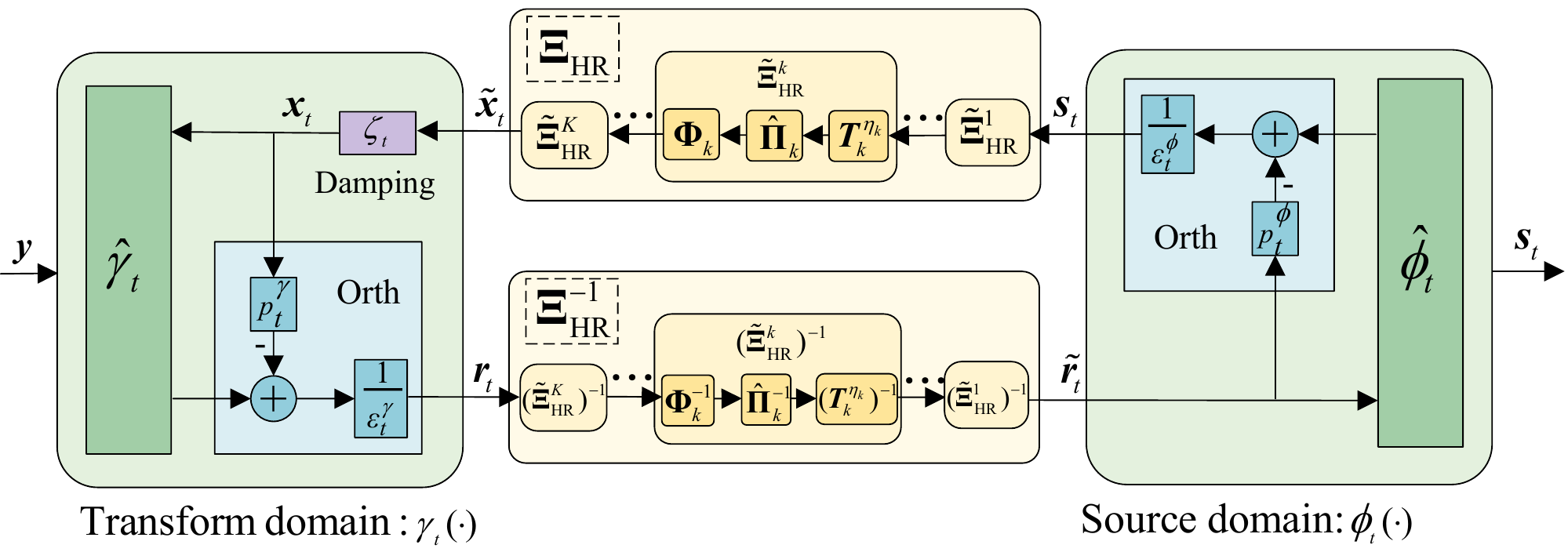}
\caption{Graphic illustration for the highly reliable CD-MAMP detector.}
\label{model3}
\end{figure}

We then develop the corresponding highly reliable CD-MAMP (HR-CD-MAMP) detector, as illustrated in Fig.~\ref{model3}. The HR-CD-MAMP still comprises the following four stages:
\begin{subequations}\label{Eqn:HR-CD-MAMP}\begin{align}
(1)&{\rm MLE}: {\bm r}_t ={\gamma}_t({\bm X}_t)=\tfrac{1}{{\varepsilon}^\gamma_t}(\bm{H}^{\mathrm H}\hat{\gamma}_t({\bm X}_t)-{\bm p}_t{\bm X}_t), \label{Eqn:CD-MAMP_a}\\
&\hat{\gamma}_t({\bm X}_t)=\theta_t{\bm B}\hat{\gamma}_{t-1}({\bm X}_{t-1})+ \xi_t({\bm y} - {\bm H}{\bm x}_t);\label{Eqn:CD-MAMP_a1}\\
(2)&{\rm HR\mbox{-}ICDT}: \tilde{{\bm r}}_t={\bm \Xi}^{-1}_{\mathrm{HR}}{\bm r}_t\label{Eqn:CD-MAMP_b}\\
&\qquad\qquad\qquad\quad\!\!\!\!\!=\prod_{k=1}^{K}\Bigl({({\boldsymbol T}_{k}^{\eta_k})}^{-1}\,\hat{\boldsymbol{\Pi}}_{k}^{-1}\boldsymbol{\Phi}_{k}^{-1}\,\Bigr){\bm r}_t;\label{Eqn:CD-MAMP_b2}\\
(3)&{\rm NLE}:{\bm s}_{t+1} ={\phi}_t(\tilde{\bm r}_t)= \tfrac{1}{{\varepsilon}^\phi_t}(\hat{\phi}_t(\tilde{\bm r}_t)-{ p}_t^\phi\tilde{\bm r}_t),\label{Eqn:CD-MAMP_c}\\
&\hat{\phi}_t(\tilde{\bm r}_t)=\mathbb{E}\{\bm s|\tilde{\bm r}_t,s_i\sim P_s\};\label{Eqn:CD-MAMP_c1}\\
(4)&{\rm HR\mbox{-}CDT}: \tilde{\bm x}_{t+1}={\bm \Xi}_{\mathrm{HR}}{\bm s}_{t+1}\label{Eqn:CD-MAMP_d}\\
&\qquad\qquad\qquad\quad\!=\prod_{k=1}^{K}\Bigl(\boldsymbol{\Phi}_{k}\,\hat{\boldsymbol{\Pi}}_{k}\,{\boldsymbol T}_{k}^{\eta_k}\Bigr){\bm s}_{t+1}.\label{Eqn:CD-MAMP_d2}
\end{align}
\end{subequations}

In the following, we present a detailed explanation of the above equations.
\begin{itemize}[leftmargin=*]
  \item {Memory linear estimator (MLE)}: In \eqref{Eqn:CD-MAMP_a}-\eqref{Eqn:CD-MAMP_a1}, the normalized parameters $\{\varepsilon^\gamma_t\}$ and orthogonal parameters $\{{\bm p}_t^{\gamma}\}$ are employed to enforce orthogonality in the iterative process. Besides, the parameters $\{\theta_t, \xi_t\}$ are optimized to improve convergence. We define $ \bm{B} = \lambda^\dagger \bm{I} - \bm{H}\bm{H}^{\mathrm H} $, and $ \lambda^\dagger = (\lambda_{\min} + \lambda_{\max})/2 $, where $\lambda_{\min}$ and $\lambda_{\max}$ are the minimal and maximal eigenvalues of $\bm{H}\bm{H}^{\mathrm H}$ \cite{MAMP}. In addition, the damping operation in Fig.~\ref{model3} is represented as $ \bm{x}_t = [\bm{X}_{t-1}, \tilde{\bm{x}}_t]\bm{\zeta}_t $, where $ \bm{\zeta}_t = [\zeta_{t,1}, \dots, \zeta_{t,t}]^T $ is a damping vector to be optimized to accelerate convergence while preserving orthogonality and Gaussianity \cite{MAMP}. The optimal $\bm{\zeta}_t$ is derived via the Lagrange multiplier method under the constraint $\sum_{i=1}^{t} \zeta_{t,i}=1$, with a closed-form solution proven to yield a monotonically non-increasing MSE sequence for guaranteed convergence, while strictly retaining the required orthogonality, as proven in Lemma 8 of \cite{MAMP}. Then the output estimate variance of the MLE is defined as:
\begin{equation}
v_{t,t}^{\gamma}=\frac{1}{N}\mathbb{E}\!\left[\|\gamma_t(\bm X_t)-\bm x\|^2\right].
\end{equation}
The estimation error $v_{t,t}^{\gamma}$ is orthogonal to those from all preceding iterations, which ensures the asymptotically IID Gaussianity of the estimation error sequence, i.e., ${\bm r}_t=\bm x+\bm z_t^\gamma$, where \(\bm z_t^\gamma\sim\mathcal{CN}(\bm 0,v_{t,t}^\gamma\bm I)\).
  \item {Highly reliable ICDT (HR-ICDT)}: In \eqref{Eqn:CD-MAMP_b}-\eqref{Eqn:CD-MAMP_b2}, we implement the ${\bm \Xi}^{-1}_{\mathrm{HR}}$ within HR-CD-MAMP to realize the high reliability of the signal detection. \(\bigl({\bm T}_{k}^{\eta_k}\bigr)^{-1}\) denotes the inverse of \({\bm T}_{k}^{\eta_k}\). For unitary transforms, the inverse coincides with the Hermitian transpose, i.e., \(\bigl({\bm T}_{k}^{\eta_k}\bigr)^{-1}=\bigl({\bm T}_{k}^{\eta_k}\bigr)^{\mathrm H}\). In particular, if \({\bm T}_{k}^{\eta_k}={\bm F}^{\mathrm H}\) or \({\bm W}^{\mathrm H}\), then \(({\bm F}^{\mathrm H})^{-1}={\bm F}\) and \(({\bm W}^{\mathrm H})^{-1}={\bm W}\). It is worth noting that each layer in ${\bm \Xi}_{\mathrm{HR}}$ consists of a phase rotation matrix $\boldsymbol{\Phi}_k$, a permutation matrix $\hat{\boldsymbol{\Pi}}_k$, and a unitary transform ${\boldsymbol T}_k^{\eta_k}$. Since all these matrices are unitary, their product is also unitary \cite{RM}. Therefore, the HR transform preserves the variance and the IID Gaussianity of the estimation error, i.e., 
\begin{equation}
\tilde v_{t,t}^{\gamma}=v_{t,t}^{\gamma}{\bm \Xi}_{\mathrm{HR}}^{\mathrm -1}({\bm \Xi}_{\mathrm{HR}}^{\mathrm -1})^{\mathrm H}=v_{t,t}^{\gamma}{\bm \Xi}_{\mathrm{HR}}^{\mathrm H}{\bm \Xi}_{\mathrm{HR}}=v_{t,t}^{\gamma}\bm I.
\end{equation}
It is worth mentioning that \(\bm\Xi_{\mathrm{HR}}\) consists of cascaded unitary and permutation operations. Therefore, it preserves the variance of the estimation error while enhancing the randomness of the signal representation, which can be written as: $\tilde{\bm r}_t=\bm s+\tilde{\bm z}_t^\gamma$, where \(\tilde{\bm z}_t^\gamma\sim\mathcal{CN}(\bm0,v_{t,t}^\gamma\bm I)\).
  \item {Non-linear estimator (NLE)}: In \eqref{Eqn:CD-MAMP_c}-\eqref{Eqn:CD-MAMP_c1}, the parameters $\{\varepsilon^\phi_t, p^\phi_t\}$ denote the normalized and orthogonal parameters in the source domain, i.e., the domain of signal $\bm s$ before applying the HR cross-domain transform ${\bm \Xi}_{\mathrm{HR}}$, analogous to those in the MLE stage. The corresponding output variance of the NLE is:
\begin{equation}
v_{t+1,t+1}^{\phi}=\frac{1}{N}\mathbb{E}\!\left[\|\phi_t(\tilde{\bm r}_t)-\bm s\|^2\right].
\end{equation}
  \item {Highly reliable CDT (HR-CDT)}: In \eqref{Eqn:CD-MAMP_d}-\eqref{Eqn:CD-MAMP_d2}, we employ the HR-CDT in the source domain to map signals into the transform domain via ${\bm \Xi}_{\mathrm{HR}}$, to achieve highly reliable signal processing, and the corresponding variance \(\tilde v_{t+1,t+1}^\phi\) is denoted as:
\begin{equation}
\tilde v_{t+1,t+1}^{\phi} = v_{t+1,t+1}^{\phi}{\bm \Xi}_{\mathrm{HR}}{\bm \Xi}_{\mathrm{HR}}^{\mathrm H}= v_{t+1,t+1}^{\phi}\bm I.
\end{equation}
\end{itemize}

In the proposed HR-CD-MAMP detector, the IMCT introduces a $K$-layer cascaded unitary transform structure. With each layer using a fast unitary transform of $\mathcal{O}(N\log N)$ complexity, the transform-related complexity scales nearly linearly with $K$. Therefore, the overall per-iteration complexity becomes $\mathcal O(PN\mathcal{T} + 2KN\mathcal{T}\log N)$, where $P$ denotes the channel taps and $T$ denotes the number of iterations. Despite this increase, the complexity growth remains moderate since $K$ is typically small in practice (such as $K=2$).
\section{Simulation Results}
This section evaluates the performance of the proposed storage-efficient interleavers, i.e., DPP and LCM interleavers, and the highly reliable transforms, i.e., IPPT and IMCT, for the enhanced RM-MAMP system. We consider a $4\times 4$ MIMO system operating over doubly-selective fading channels modeled according to 3GPP TR $38.901$. We adopt the standardized tapped delay line (TDL) channel family (including typical profiles TDL-A, TDL-C, and TDL-D), each of which characterizes realistic multipath propagation via discrete taps with standardized delay offsets and power levels. To ensure a fair comparison, all multiplexing schemes are configured to occupy the same total bandwidth. Specifically, OTFS employs a subcarrier spacing of $\Delta f = 15$ kHz with an OTFS multiplexing matrix ${\bm F}_Q^{\mathrm H}\!\otimes{\bm I}_P$, whereas IFDM, AFDM, and OFDM use a reduced subcarrier spacing of $\Delta f/Q$. QPSK modulation is adopted throughout. Then, we perform iterative detection with CD-MAMP/OAMP, and the signal-to-noise ratio (SNR) is defined as ${\rm SNR}=1/\sigma^2$.
\begin{figure}[htbp]
\centering
 \includegraphics[width=0.9\linewidth]{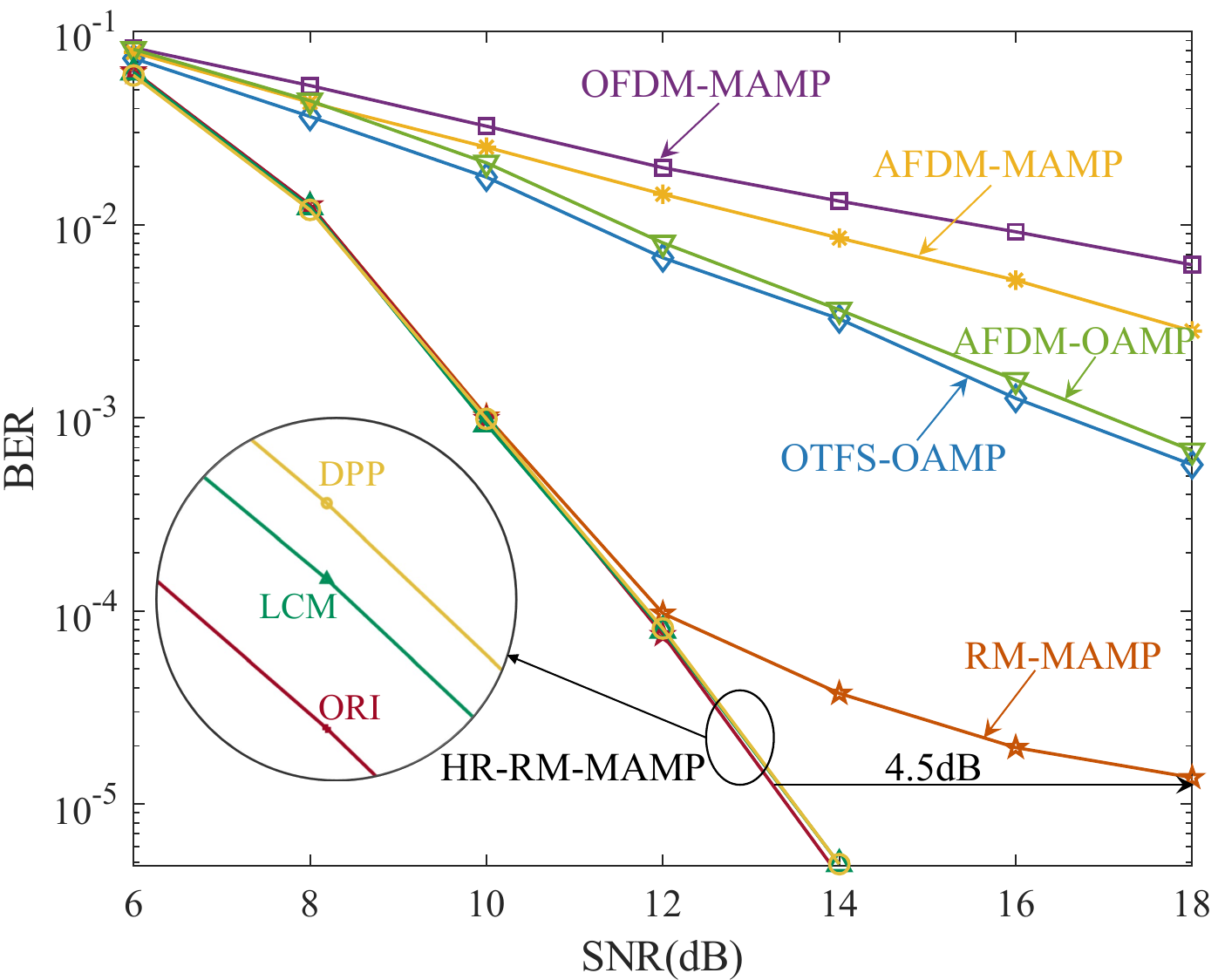}
\caption{A comparison of storage-efficient design methods in RM-MAMP and other multicarrier communication systems.}
\label{result1}
\end{figure}

First, we compare bit-error-rate (BER) versus SNR performance of the proposed RM-MAMP with OFDM, OTFS, and AFDM benchmarks, all equipped with the state-of-the-art CD-OAMP/CD-MAMP detectors \cite{IFDM,CD-OAMP,AFDM-MAMP,RM}, abbreviated as OFDM/OTFS/AFDM-OAMP/MAMP. All schemes use 256 subcarriers, with a 100 km/h device speed to introduce Doppler spread. As shown in Fig.~\ref{result1}, RM-MAMP outperforms the other schemes by about $7$ dB, confirming its advantage in this setting \cite{IFDM}. However, under the ill-conditioned TDL-A channel with high mobility, the RM-MAMP system exhibits a noticeable BER error floor, in contrast to its performance under typical time-varying multipath channels. We then employ the IPPT-OUT in Section IV as the representative scheme for highly reliable RM (HR-RM) with CD-MAMP (denoted as HR-RM-MAMP), while detailed comparisons among all highly reliable schemes are provided in the following experiment. Fig.~\ref{result1} shows that the HR-RM-MAMP achieves about a $4.5$ dB improvement over the RM-MAMP at nearly identical BER. This indicates that the HR-RM-MAMP enhances robustness in severely ill-conditioned channels by mitigating the degradation observed in RM-MAMP. In addition, when the original (ORI) interleaver in HR-RM is replaced by the DPP or LCM interleaver, the BER curves overlap closely with the ORI curve, indicating that our storage-efficient interleavers achieve near-identical BER performance to the ORI interleaver for HR-RM-MAMP.

\begin{figure}[htbp]
\centering
\includegraphics[width=0.9\linewidth]{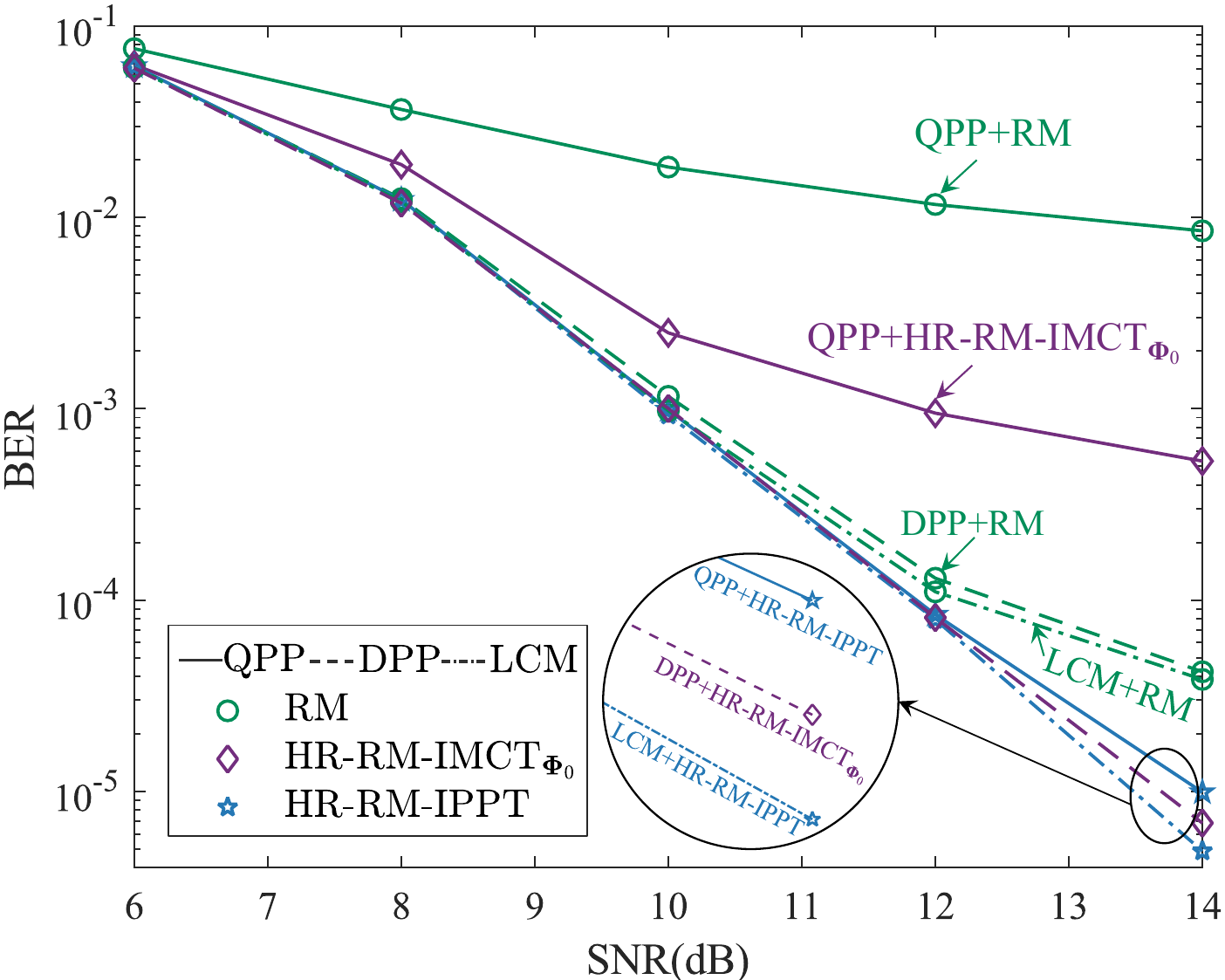}
\caption{A performance comparison of QPP, LCM, and DPP interleavers in RM-MAMP and HR-RM-MAMP systems.}
\label{result1_2}
\end{figure}

Fig.~\ref{result1_2} compares the BER performance of the proposed storage-efficient interleavers with a conventional structured quadratic permutation polynomial (QPP) interleaver under both RM-MAMP and HR-RM-MAMP systems. Specifically, QPP interleavers are widely regarded as a practical standard in Long Term Evolution (LTE) and turbo-coded systems owing to their low-complexity, storage-efficient, and deterministic construction \cite{QPP,QPP-2}. Therefore, QPP provides a representative benchmark of conventional structured interleavers, against which the superiority of the proposed interleavers can be clearly demonstrated. In the HR-RM configuration, the IMCT is employed with $K=2$ and $\boldsymbol{\Phi}_k=\boldsymbol{I}$, which is sufficient to compensate for the performance degradation of the original RM system, as discussed in Section IV-2. For clarity, this configuration is referred to as $\mathrm{IMCT}_{\boldsymbol{\Phi}_0}$. In the baseline RM-MAMP system without high-reliability enhancement, the QPP interleaver exhibits a pronounced BER loss compared with the proposed LCM and DPP interleavers, while LCM and DPP achieve nearly identical performance. This gap stems from the inherent deterministic structure of QPP, which restricts the asymptotic randomness of the permutation and fails to meet the core constraints of the universality class in RM-MAMP required for optimal detection performance. When $\mathrm{IMCT}_{\boldsymbol{\Phi}_0}$ is applied, the performance degradation of QPP remains significant, whereas the DPP interleaver achieves a substantial performance improvement. With the introduction of phase perturbation through the IPPT-based HR-RM-MAMP system, the performance gap between QPP and LCM is reduced, although a BER loss persists for QPP. These results demonstrate the effectiveness of the proposed storage-efficient interleavers and highlight the necessity of combining them with highly reliable transforms to achieve robust performance in challenging channel conditions.

\begin{figure}[htbp]
\centering
\includegraphics[width=0.95\linewidth]{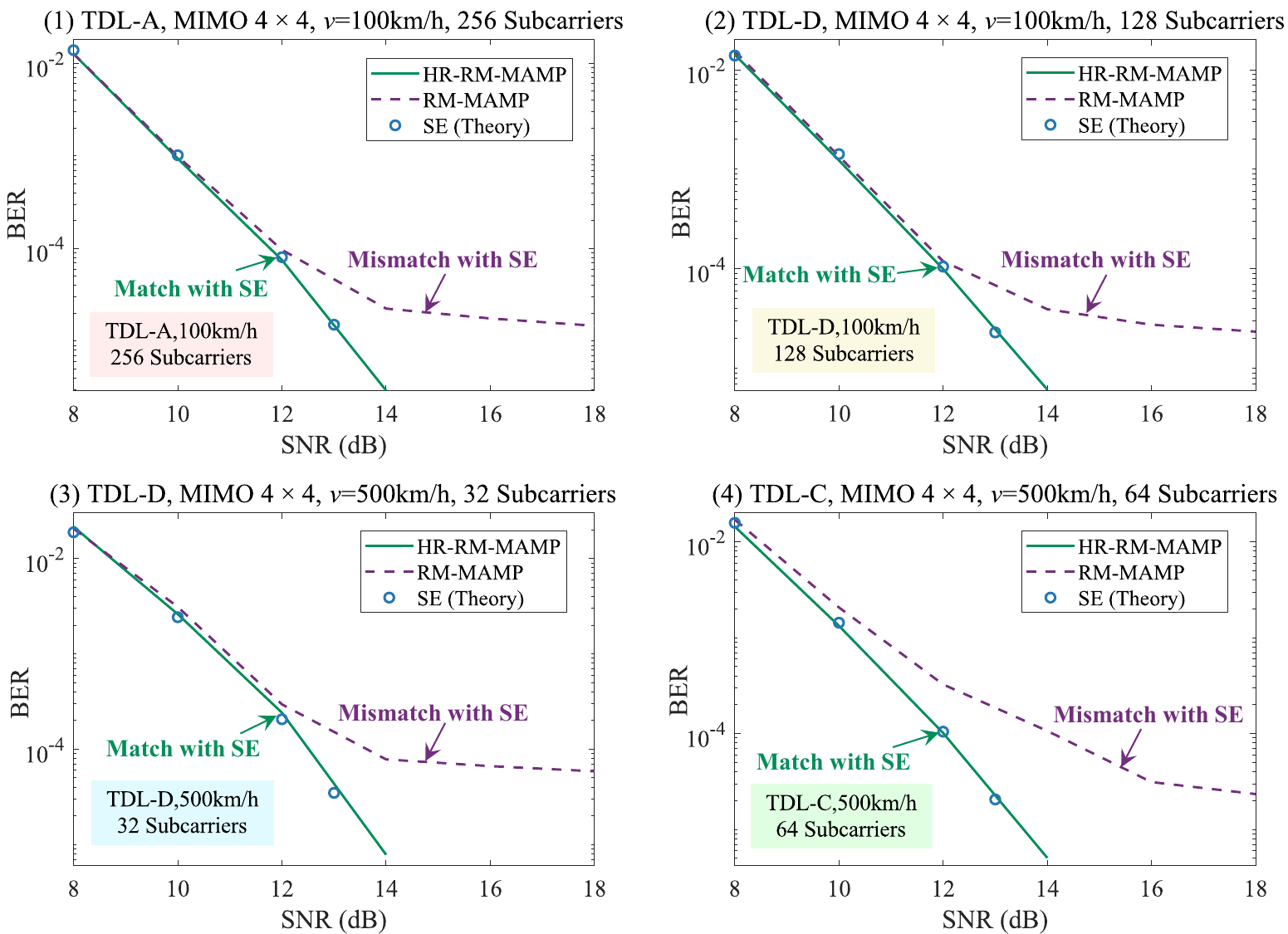}
\caption{A BER comparison of HR-RM-MAMP and RM-MAMP with DPP interleaver under different channel conditions.}
\label{multipic}
\end{figure}

\begin{figure}[htbp]
\centering
\includegraphics[width=0.9\linewidth]{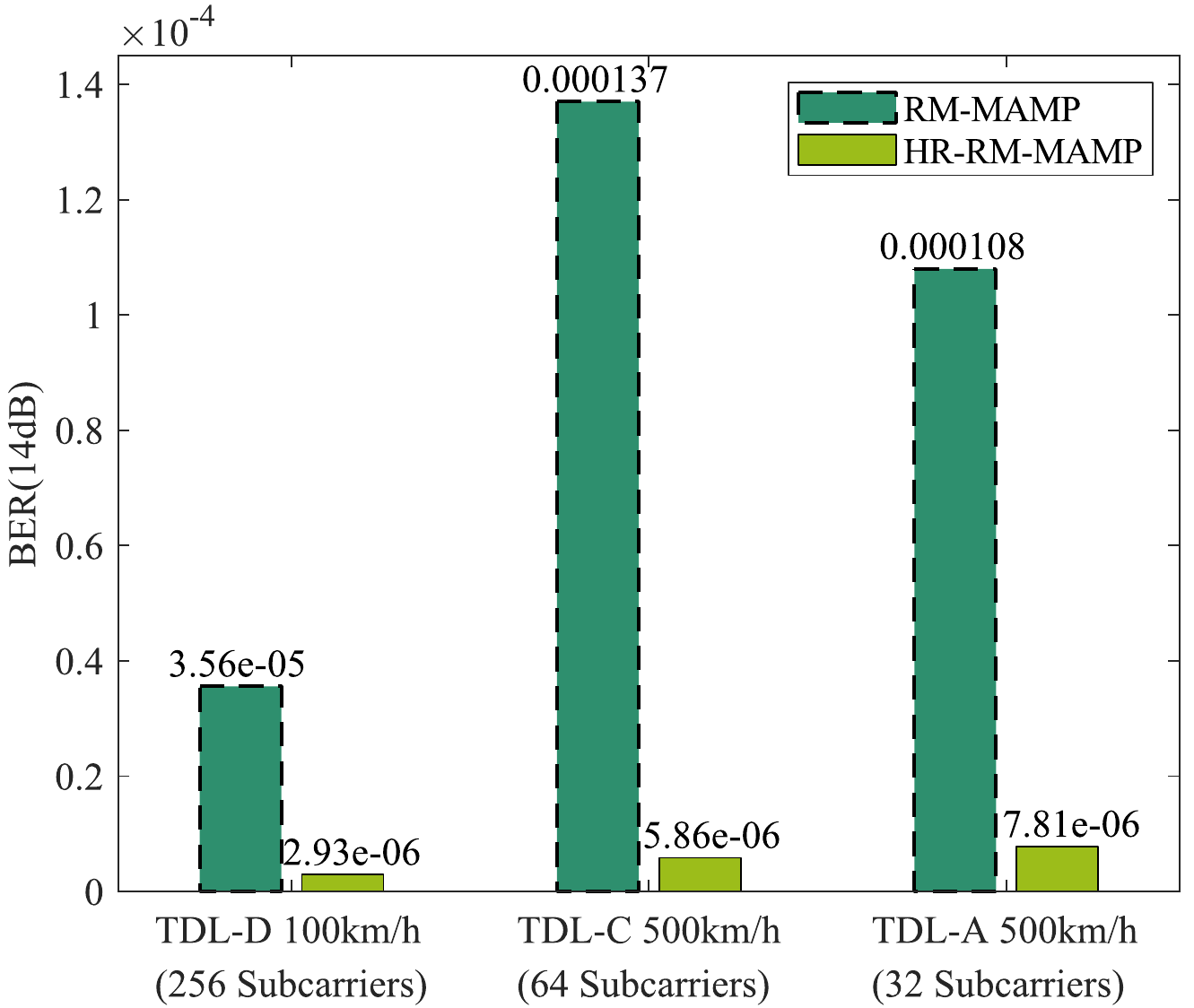}
\caption{A BER comparison at a fixed SNR of HR-RM-MAMP and RM-MAMP under different channel conditions.}
\label{result2}
\end{figure}

We further evaluate the performance of the proposed HR-RM-MAMP under diverse realistic channel conditions. The evaluation covers three standardized 3GPP TDL channel models (TDL-A, TDL-C, TDL-D), two mobility scenarios ($100$ km/h and $500$ km/h), and four system scales with $32$, $64$, $128$, and $256$ subcarriers, to fully validate the robustness of the proposed design in both typical and severely ill-conditioned propagation environments. All comparison schemes adopt the proposed DPP interleaver to ensure fair performance evaluation. Fig.~\ref{multipic} compares the BER performance against SNR for HR-RM-MAMP and RM-MAMP, alongside the theoretical performance bound derived from SE. The results show that only the proposed HR-RM-MAMP achieves near-perfect alignment between the simulated BER curve and the SE bound across all tested channel models, mobility levels, and system scales. In contrast, RM-MAMP exhibits a clear mismatch with the SE bound, with significant BER performance degradation compared to HR-RM-MAMP. To further quantify the performance gain, Fig.~\ref{result2} compares the BER of HR-RM-MAMP and conventional RM-MAMP at a fixed SNR of 14 dB across all evaluated configurations. The results demonstrate that HR-RM-MAMP consistently outperforms RM-MAMP under varying channel types, mobility levels, and system scales. In particular, for small-scale systems with severe channel ill-conditioning, RM-MAMP suffers from severe performance collapse, while HR-RM-MAMP maintains strong robustness. This validates the effectiveness of the proposed highly reliable transform design in enhancing RM-MAMP for \textit{finite and severely ill-conditioned channels}. It is noted that in the extremely small-scale scenario with $12$ subcarriers, both schemes exhibit noticeable BER degradation due to the large-system limit dependency of AMP-type algorithms. HR-RM-MAMP still retains a substantial performance advantage over the RM-MAMP baseline, and the IBST framework mentioned in Section II can be adopted for further optimization of our proposed HR-RM-MAMP in such scenarios \cite{IBST}.


\begin{figure}[htbp]
\centering
 \includegraphics[width=0.9\linewidth]{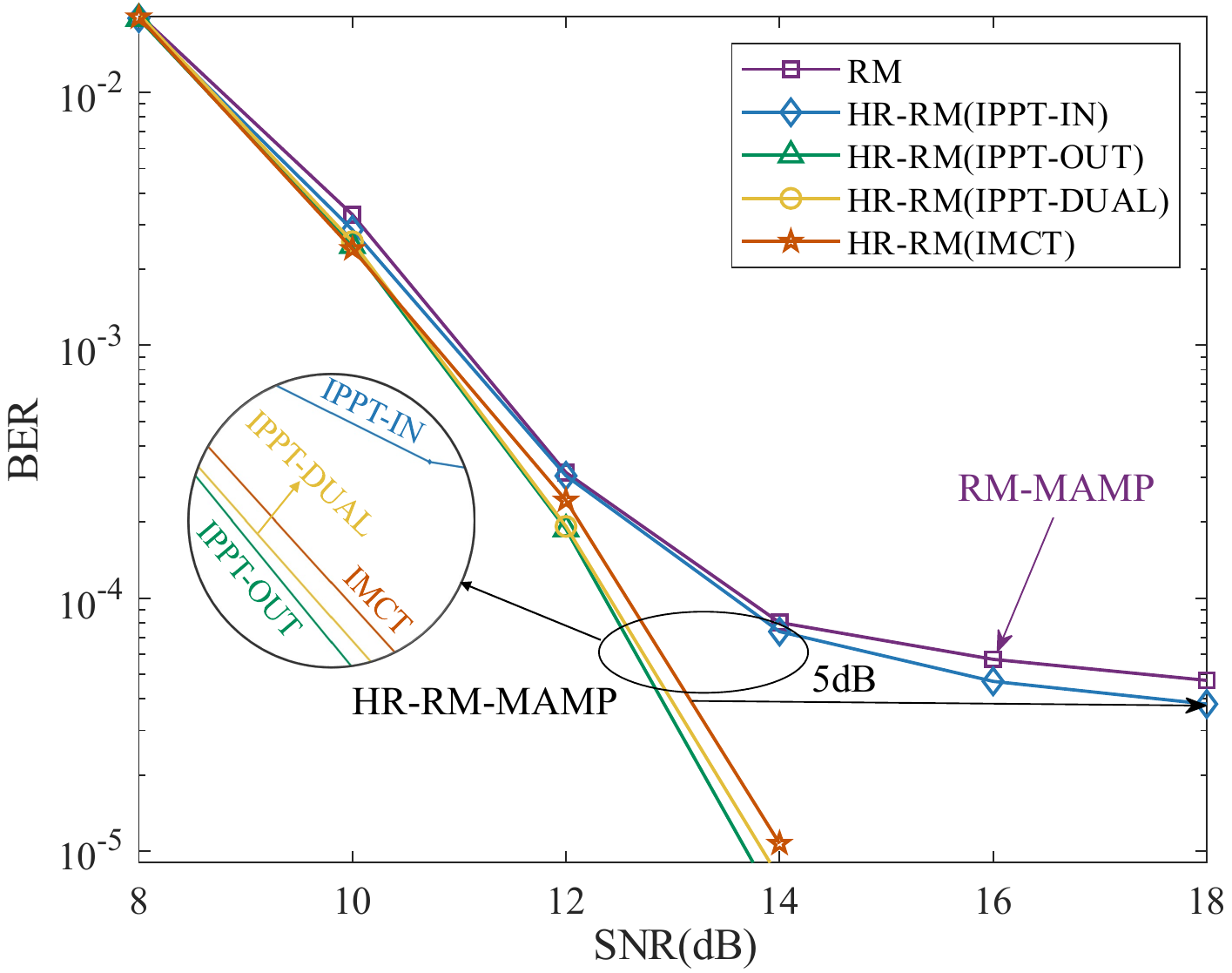}
\caption{A comparison of original RM-MAMP and HR-RM-MAMP.}
\label{result3}
\end{figure}

Then, we consider a smaller $4\times4$ MIMO system with $32$ subcarriers to compare all highly reliable schemes, i.e., IPPT-IN, IPPT-OUT, IPPT-DUAL, and IMCT in Section IV. The results in Fig.~\ref{result3} show that HR-RM with IPPT-IN exhibits no noticeable improvement with CD-MAMP, whereas both IPPT‐OUT and IPPT‐DUAL achieve gains exceeding $5$ dB relative to the baseline RM-MAMP. It is worth noting that the proposed IMCT scheme achieves a comparable BER improvement even with only a dual-layer coupling structure and without any phase perturbation. However, in scenarios where IPPT-based schemes become less effective, extending to multi-layer interleaved coupling can still be considered to further enhance robustness. Since IPPT-OUT and IPPT-DUAL perform equivalently, but IPPT-OUT is simpler to implement, IPPT-OUT is selected as the representative highly reliable design scheme. For simplicity, we refer to IPPT-OUT as IPPT unless otherwise specified.

\begin{figure}[htbp]
\centering
 \includegraphics[width=0.9\linewidth]{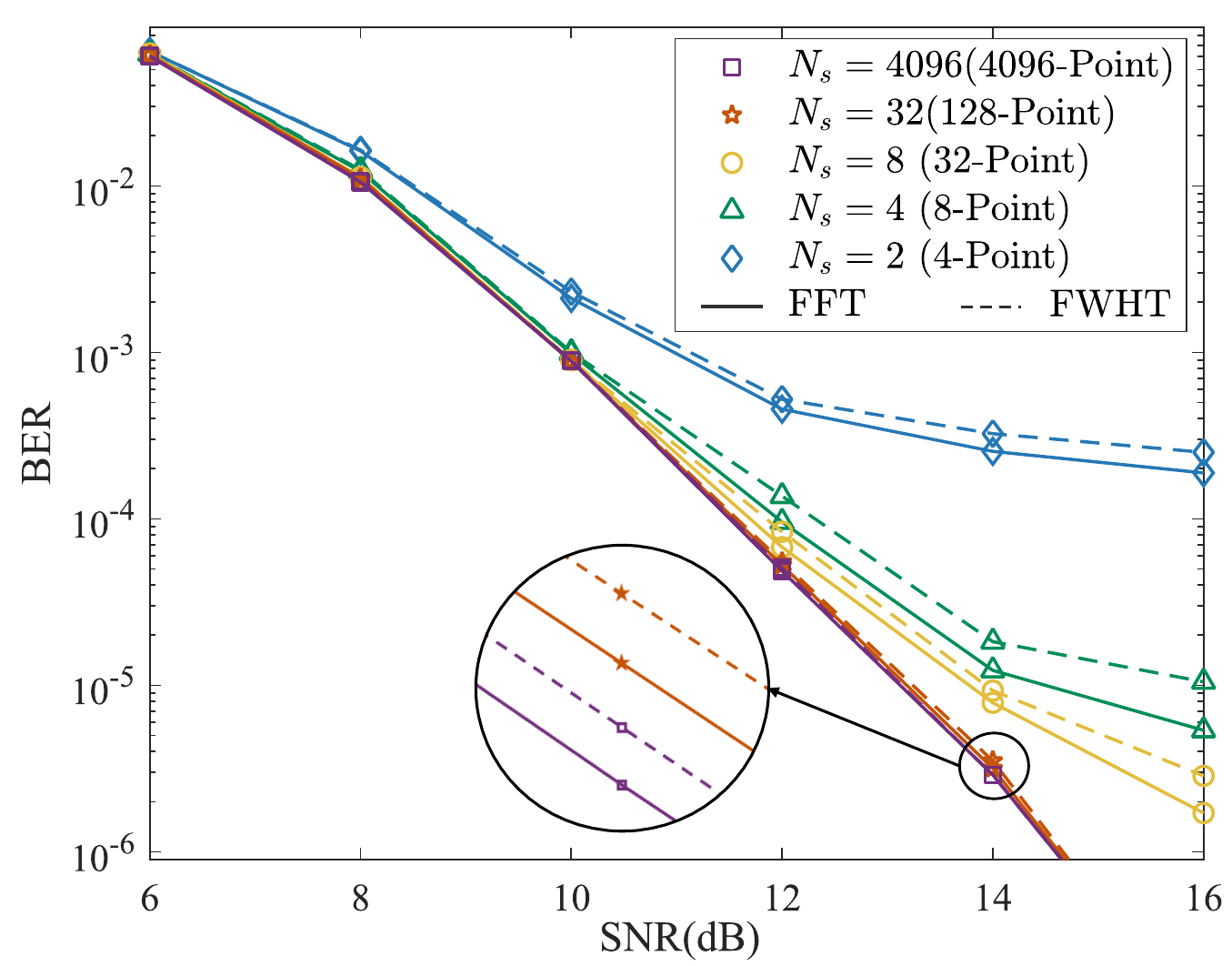}
\caption{A comparison of IBS-HR-RM and HR-RM, both integrated with CD-MAMP, based on different transform bases.}
\label{result4}
\end{figure}

Fig.~\ref{result4} shows the BER performance of the highly reliable IBS-RM-MAMP at different implementation scales $N_s$ in a $4\times4$ MIMO system with $1024$ subcarriers. Here, $L=N/N_s$ denotes the number of separated blocks. When $N_s=32$ for IPPT-based highly reliable IBS-RM (IBS-HR-RM), the BER curve coincides with that of the full-scale HR-RM ($N_s=4096$), both employing CD-MAMP. This shows that the implementation scale can be reduced to $N_s=32$ without degrading the BER performance. As $N_s$ decreases further, BER degrades for FFT-based and FWHT-based implementations, and at each reduced $N_s$, the FFT-based design outperforms the FWHT-based design, whereas FWHT offers lower hardware implementation complexity and greater resource efficiency. These results confirm that IBS-HR-RM-MAMP allows for a much smaller implementation scale while preserving full-scale BER performance.

\begin{figure}[htbp]
\centering
 \includegraphics[width=0.9\linewidth]{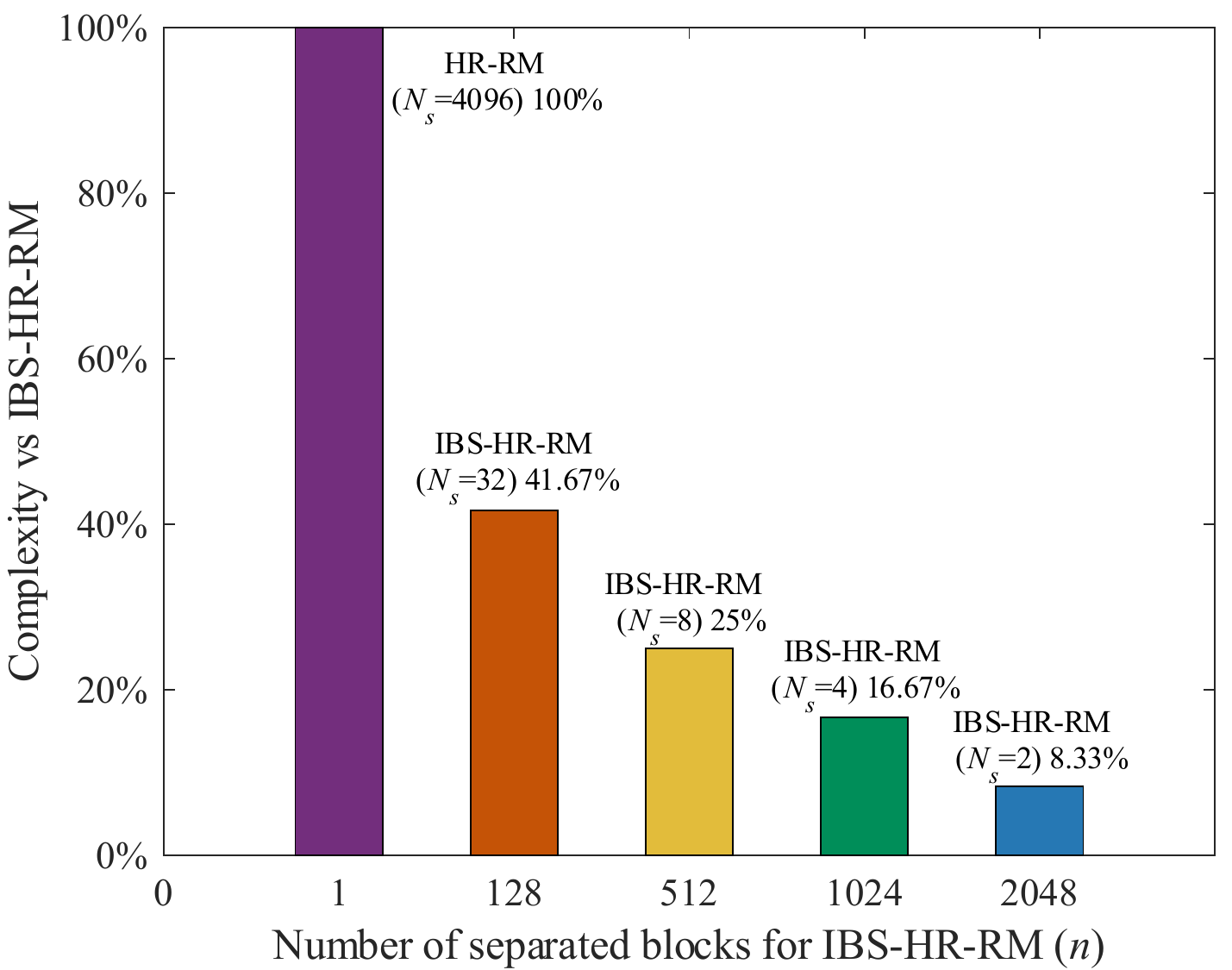}
\caption{A complexity comparison of IBS-HR-RM and HR-RM, both integrated with CD-MAMP.}
\label{result5}
\end{figure}

Fig.~\ref{result5} presents the \emph{relative complexity} $\Theta_{\rm IBS}$ versus $N_s$, with $ \Theta_{\rm IBS} = {\cal O}(\log_{N}{N_s})={\cal O}(1-\log_{N}{L})$ \cite{IBST}. As $N_s$ is reduced from $4096$ to $32$, $\Theta_{\rm IBS}$ falls by more than $50\%$ and the implementation scale is reduced to less than $1\%$, while the BER remains unchanged at $N_s=32$ (see Fig.~\ref{result4}). This demonstrates a clear trade-off, i.e., by adjusting $N_s$ and choosing FFT or FWHT as the transform basis, the implementation complexity can be substantially lowered with minimal impact on BER performance, enabling flexible design for IBS-HR-RM with CD-MAMP. To quantitatively characterize the complexity-BER tradeoff, we tabulate the measured metrics across different $N_s$ in Table~\ref{tab:tradeoff}. The table confirms that IBS-HR-RM with $N_s=32$ retains near-identical BER performance to full-scale HR-RM while reducing relative complexity to $41.67\%$, consistent with Fig.~\ref{result4}. As $N_s$ decreases further, the implementation complexity falls continuously with gradual BER degradation. This quantified tradeoff offers a clear design guideline: $N_s$ can be flexibly tuned to balance implementation overhead and detection reliability for practical hardware deployment.

\begin{table}[]
\centering
\caption{Quantitative Complexity-BER Tradeoff of IBS-HR-RM.}
\label{tab:tradeoff}
\scalebox{0.78}{\begin{tabular}{|c|c|c|c|c|}
\hline
Scheme & $N_s$ & \begin{tabular}[c]{@{}c@{}}Relative \\ Complexity $\Theta_{\rm IBS}$\end{tabular} & \begin{tabular}[c]{@{}c@{}}BER (14dB)\\ FFT-based\end{tabular} & \begin{tabular}[c]{@{}c@{}}BER (14dB)\\ FWHT-based\end{tabular} \\ \hline
HR-RM & 4096 & 100\% & $3.02\times 10^{-6}$ & $3.21\times 10^{-6}$ \\ \hline
\multirow{4}{*}{IBS-HR-RM} & 32 & 41.67\% & $3.17\times 10^{-6}$ & $3.52\times 10^{-6}$ \\ \cline{2-5} 
 & 8 & 25\% & $7.86\times 10^{-6}$ & $9.30\times 10^{-6}$ \\ \cline{2-5} 
 & 4 & 16.67\% & $1.23\times 10^{-5}$ & $1.83\times 10^{-5}$ \\ \cline{2-5} 
 & 2 & 8.33\% & $2.54\times 10^{-4}$ & $3.25\times 10^{-4}$ \\ \hline
\end{tabular}}
\end{table}

\begin{figure}[htbp]
\centering
 \includegraphics[width=0.9\linewidth]{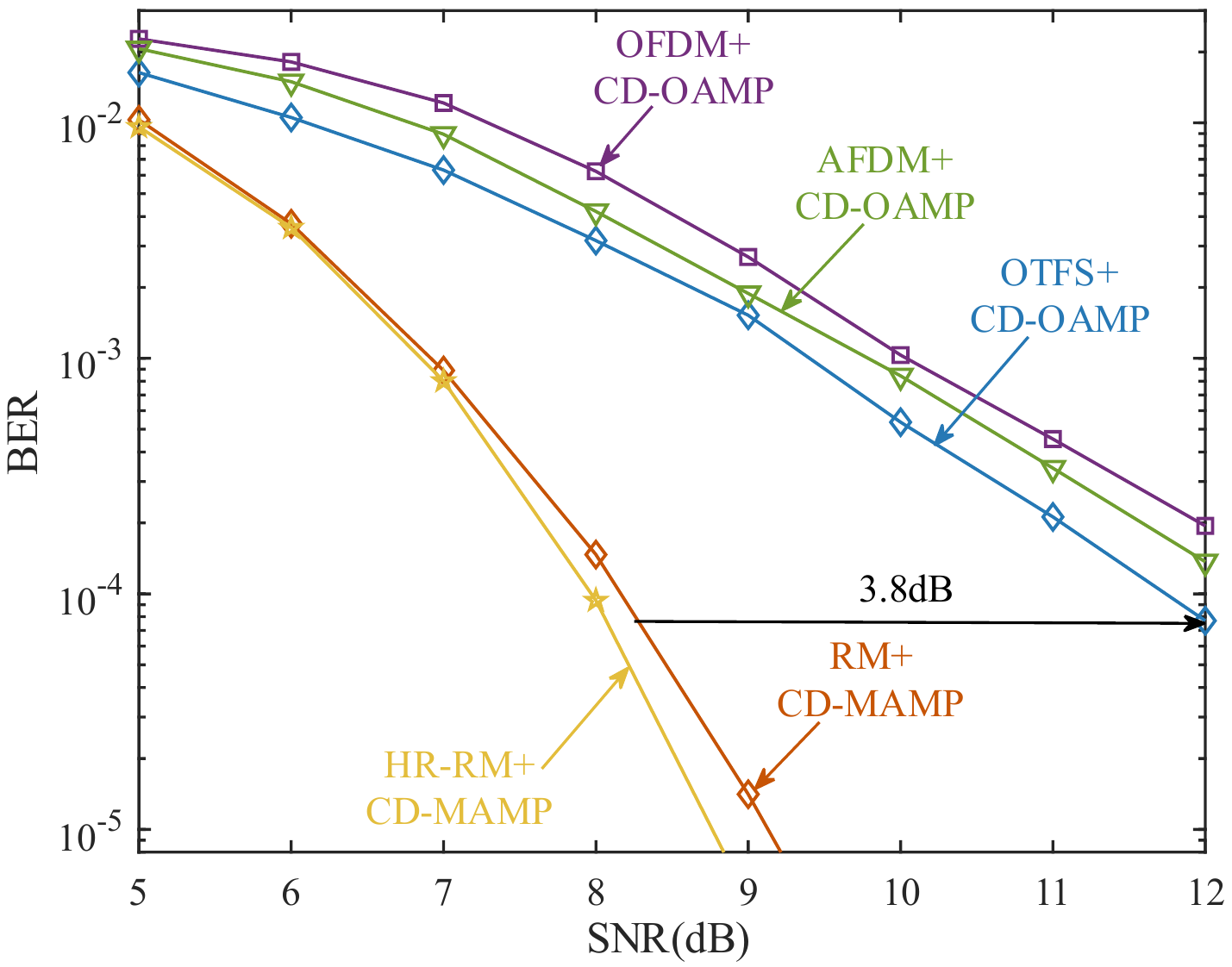}
\caption{A comparison of coded HR-RM-MAMP and other coded multicarrier communication systems.}
\label{result6}
\end{figure}

Fig.~\ref{result6} presents the BER performance of the low-density parity-check (LDPC)-coded RM-MAMP system ($\mathrm {Code Length=2048}$, $\mathrm{Code Rate=0.625}$) in the MIMO $4\times4$ with $1024$ subcarriers under a $100$ km/h TDL-A channel. The RM-MAMP outperforms OFDM, OTFS, and AFDM with OAMP by at least $3.8$ dB, confirming its advantage in the coded scenario. These results demonstrate that RM-MAMP retains its superiority in the presence of LDPC coding and that HR-RM with CD-MAMP offers further reliability improvement in the coded system. 

\section{Conclusion}
This paper presents a storage-efficient and highly reliable design for the RM-MAMP system, aiming to address two limitations that hinder its practical deployment: the storage overhead of conventional interleavers and the performance degradation in {\it finite and severely ill-conditioned channels}. To mitigate the high memory consumption and transceiver signaling overhead of conventional random interleavers, we propose two storage-efficient interleavers, namely a Logistic chaotic mapping interleaver with a quantitative parameter-selection criterion and a dual-stage high-order permutation polynomial interleaver. Both schemes reduce the interleaver storage complexity from $\mathcal{O}(N)$ to $\mathcal{O}(1)$, while achieving near-identical BER performance to fully random interleavers with drastically reduced storage requirements. To improve system reliability under severely time-varying conditions, we further develop a highly reliable interleaved transform framework that incorporates interleaved phase perturbation and multi-layer interleaved coupling, which improve the incoherence and diversity of the equivalent channel matrix. Simulation results verify that the proposed interleavers achieve nearly identical BER as conventional designs, and the highly reliable transforms provide more than $4$ dB performance gains in {\it finite and severely ill-conditioned channels}, demonstrating both storage efficiency and high reliability for enhanced RM-MAMP in practical scenarios. Overall, this work offers an efficient and reliable solution for enabling an enhanced RM-MAMP system in future 6G wireless communications.
\section*{Acknowledgment}{The work was supported by the National Natural Science Foundation of China (Grant Nos. 62371428, 62301485, 62394292, 62571399, 62471357), the Zhejiang Provincial Natural Science Foundation (Grant No. LZ25F010002), and the A*STAR under the RIE2025 Industry Alignment Fund–Industry Collaboration Projects (IAF-ICP) Funding Initiative (Award No. I2501E0045), as well as cash and in-kind contributions from the industry partner(s).}
%%@@@@@@@@@@@@@@@@@@@@@@@@@@@@@@@@@@@@@@@@@@@@
\bibliographystyle{IEEEtran}
\bibliography{refs}

\newpage
\begin{IEEEbiography}[{\includegraphics[width=1in,height=1.25in,clip,keepaspectratio]{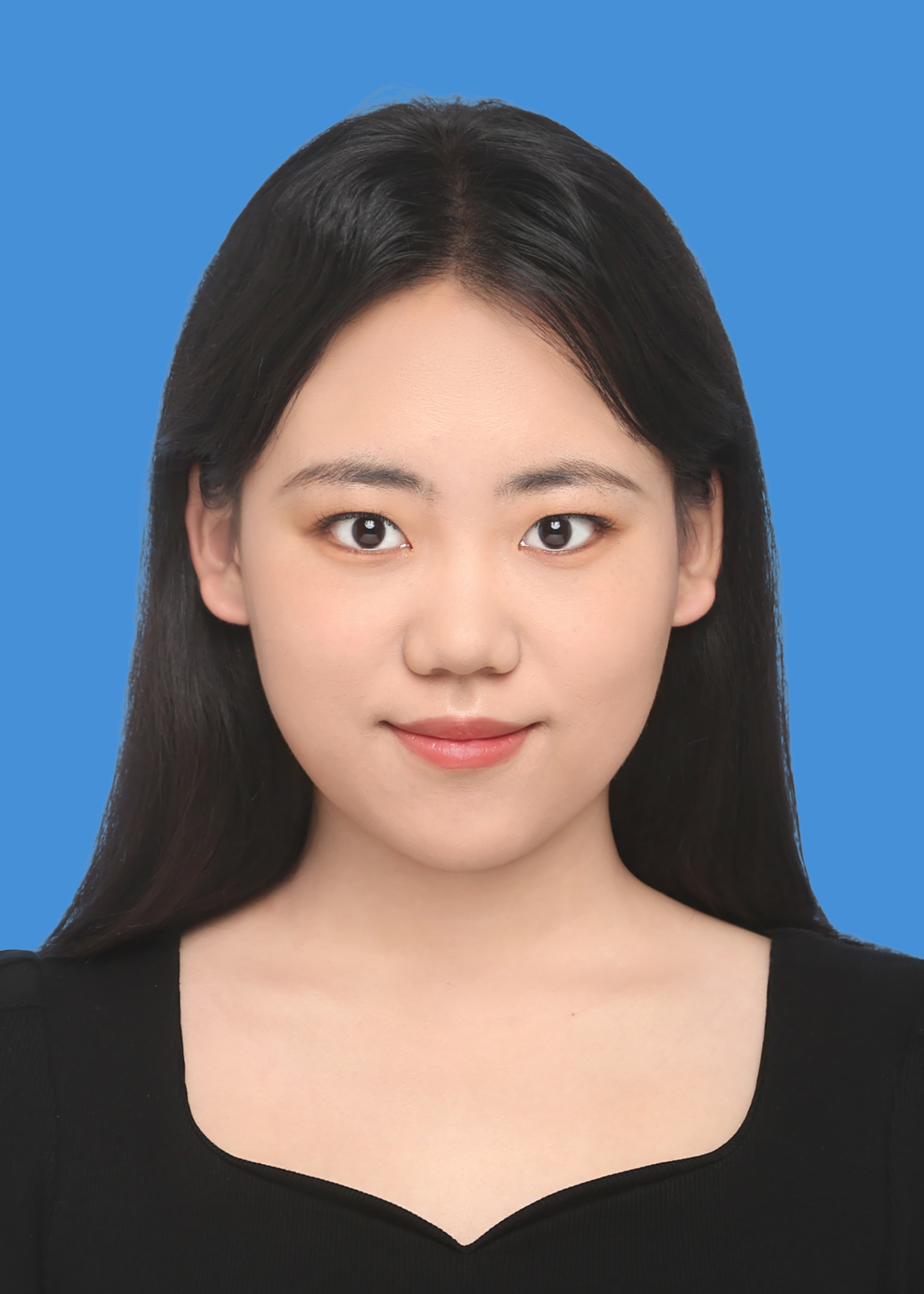}}]{Ming Wang}
was born in Lanzhou, Gansu, China. She received the B.S. degree in 2023 from the Communication University of China, Beijing, China. She is currently working toward the M.S. degree at the School of Information and Communication Engineering, Communication University of China. Her research interests include wireless communication, approximate message passing, and signal detection.
\end{IEEEbiography}

\begin{IEEEbiography}[{\includegraphics[width=1in,height=1.25in,clip,keepaspectratio]{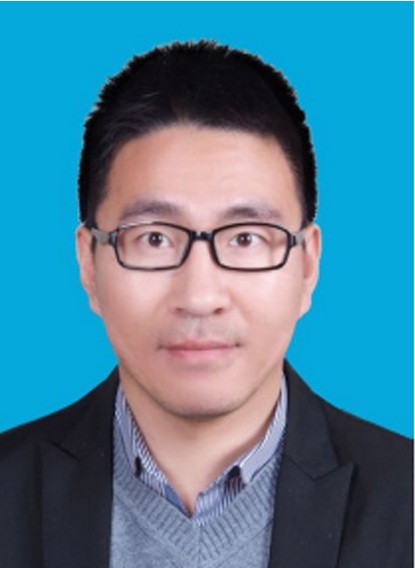}}]{Shufeng Li, Member, IEEE}
(Member, IEEE) received the B.Sc. and M.Sc. degrees from Hebei University of Technology, Tianjin, China, in 2004 and 2007, and Ph.D. degree from Beihang University, Beijing, China, in 2011. He is currently a professor with the School of Information and Engineering, Communication University of China, Beijing, China. His research interests mainly include channel estimation, semantic communication, massive MIMO communication, and non-orthogonal multiple access.
\end{IEEEbiography}

\begin{IEEEbiography}[{\includegraphics[width=1in,height=1.25in,clip,keepaspectratio]{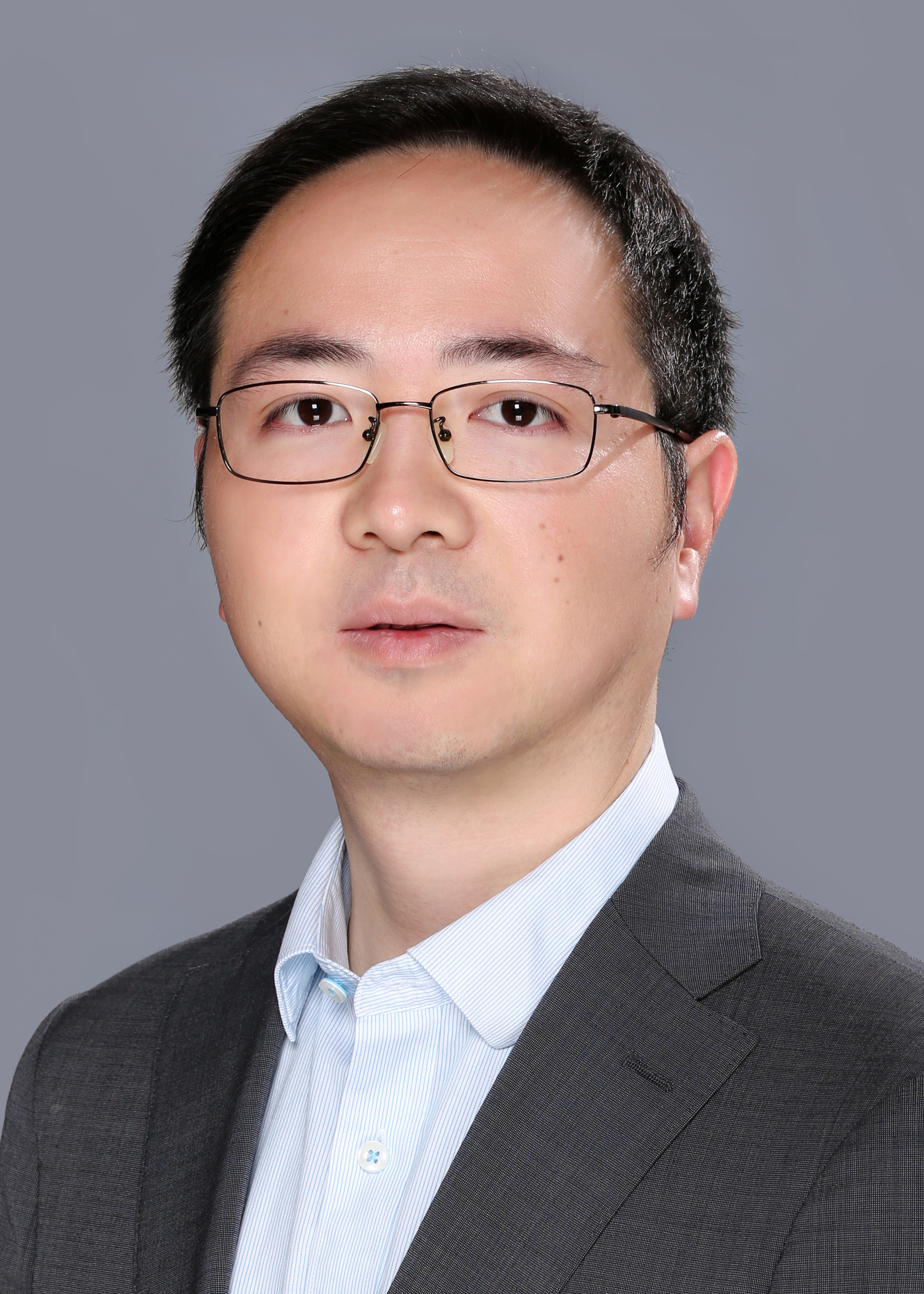}}]{Lei Liu, Senior Member, IEEE}
(Senior Member, IEEE) received the Ph.D. degree in Communication and Information Systems from Xidian University in 2017. From 2014 to 2016, he was an exchange Ph.D. student with Nanyang Technological University (NTU). From 2016 to 2017, he was a Post-Doctoral Research Fellow with the Singapore University of Technology and Design (SUTD). From 2017 to 2019, he was a Postdoctoral Fellow with the City University of Hong Kong (CityU), where he was a Research Fellow in 2019. From 2019 to 2023, he was an Assistant Professor with the Japan Advanced Institute of Science and Technology (JAIST). He is currently a Tenure-Track ZJU Young Professor with Zhejiang University. His research interests cover a broad range of message passing, wireless communications, machine learning, statistical signal processing, modern channel coding, and information theory.

Dr. Liu was granted the National Natural Science Fund for the Excellent Young Scientists (Overseas) by NSFC in 2023, and was named the ``6G New Horizon Young Scientist” at Global 6G Conference 2025. He received the Young Star Award and two Best Poster Awards at the Chinese Institute of Electronics Conference on Information Theory (CIEIT), as well as the Best Paper Award at IEEE PIMRC 2025. He served as a Youth Committee Member in the Signal Processing and Communications Societies of CIE, an Editor for \textsc{Entropy}, a Youth Editor for the \textsc{Chinese Journal of Electronics}, \textsc{Journal of Information and Intelligence}, \textsc{Electronics and Signal Processing}, and the Lead Guest Editor for \textsc{Entropy} Special Issue on Advances in New Physical Layer Technologies for Next-Generation Wireless Communications, the co-organizer of ISIT 2026 Workshop on Next-Generation Waveforms Design for Communications, Sensing, and Integrated Systems, the Publications Co-Chair of the 2021 IEEE ITW, TPC Co-Chair for NCIC 2023 and ICCTIT 2023, Session Co-Chair for WCSP 2023, etc.
\end{IEEEbiography}

\begin{IEEEbiography}[{\includegraphics[width=1in,height=1.25in,clip,keepaspectratio]{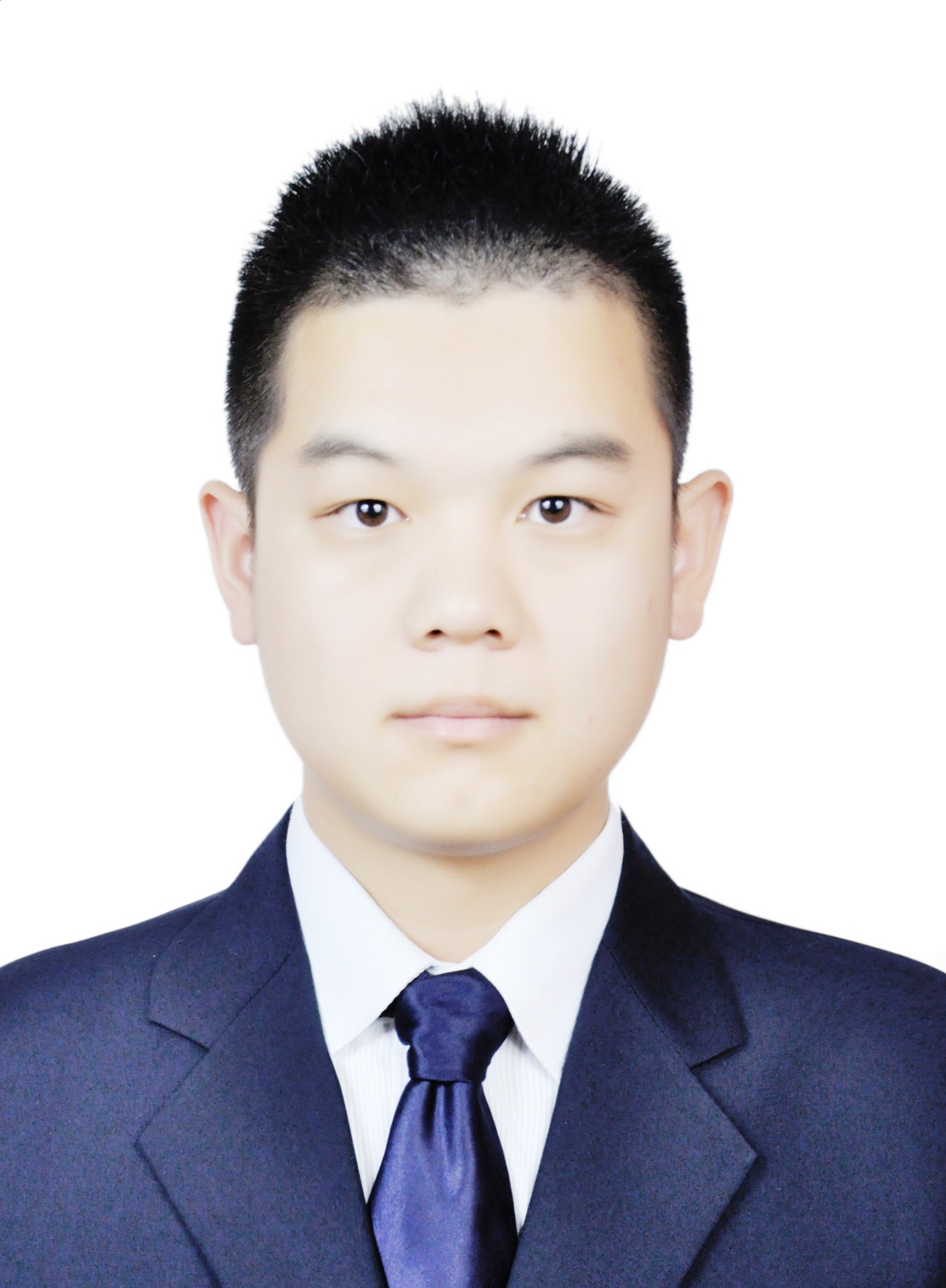}}]{Yao Ge, Member, IEEE}
(Member, IEEE) received the Ph.D. degree in Electronic Engineering from The Chinese University of Hong Kong (CUHK), Shatin, Hong Kong, in 2021, and the M.Eng. degree (research) in Communication and Information System and the B.Eng. degree in Electronics and Information Engineering from Northwestern Polytechnical University (NPU), Xi’an, China, in 2016 and 2013, respectively. He is currently a Research Fellow with the AUMOVIO-NTU Corporate Lab, Nanyang Technological University (NTU), Singapore. From October 2015 to March 2016, he was a Visiting Scholar with the Department of Electrical and Computer Systems Engineering, Monash University, Melbourne, VIC, Australia. From April 2016 to August 2016, he was a Visiting Scholar with the Department of Computer, Electrical and Mathematical Science and Engineering, King Abdullah University of Science and Technology (KAUST), Thuwal, Saudi Arabia. From May 2019 to December 2019, he was a Visiting Scholar with the Department of Electrical and Computer Engineering, University of California at Davis (UC Davis), Davis, CA, USA. His current research interests include wireless communications and system design, the Internet of Things, cognitive radio networks, automotive vehicle signal processing and communications, integrated sensing and communications, wireless network security, statistical signal processing, optimization, and game theory. He received the Best Paper Award from the International Conference on Wireless Communications and Signal Processing (WCSP) 2022 and the Best Poster Award of the Chinese Institute of Electronics Conference on Information Theory 2024. He is a Founding Member of the IEEE ComSoc Special Interest Group (SIG) on OTFS and has served as the Organizer/Chair of workshops and special sessions in numerous international conferences. He is also a Youth Editor of the Engineering, Journal of Information and Intelligence and a Regional Editor of the Computing $\&$ AI Connect.
\end{IEEEbiography}

\begin{IEEEbiography}[{\includegraphics[width=1in,height=1.25in,clip,keepaspectratio]{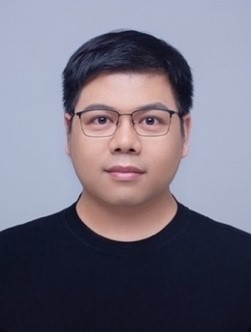}}]{Yuhao Chi, Senior Member, IEEE}
(Senior Member, IEEE) was born in Yingxian, Shanxi, China. He received the Ph.D. degree in communication and information systems from Xidian University (XDU), Xi’an, China, in 2018. 

From 2016 to 2017, he was an Exchange Ph.D. Student with Nanyang Technological University (NTU), Singapore, supported by the State Scholarship Fund from China Scholarship Council, and a Visiting Student with Singapore University of Technology and Design (SUTD), Singapore. From 2018 to 2021, he was a Senior Engineer with Huawei Technologies Company Ltd. He is currently an Associate Professor at XDU. His research interests include multicarrier waveform, multiuser coding and detection, message passing algorithms, wireless communications, and deep learning. He received the Best Poster Award at the 30th and 31st Chinese Institute of Electronics Conference on Information Theory (CIEIT 2023 and 2024). He served as the Session Co-Chair for IEEE ICCT 2023 and IEEE ISIT 2026 Workshop, and as an Editor for Entropy Special Issue on ``Advances in New Physical Layer Technologies for Next-Generation Wireless Communications.”
\end{IEEEbiography}

\end{document}